\documentclass[10pt,sigconf,letterpaper,nonacm]{acmart}
\renewcommand\footnotetextcopyrightpermission[1]{}

\usepackage{tikz}
\usepackage{amsmath}

\usepackage{quoting}

\usepackage{filecontents}
\usepackage{soul}
\newcommand{\totalNumberOfApps}[0]{1767\xspace}
\newcommand{\sibapp}[0]{Store A\xspace}
\newcommand{\iapps}[0]{Store B\xspace}
\newcommand{\appstar}[0]{Store C\xspace}

\newcommand{\sibappIranianPercentage}[0]{45\%\xspace}

\newcommand{\sibappTotal}[0]{558}
\newcommand{\iappsTotal}[0]{1323\xspace}
\newcommand{\appstarTotal}[0]{221\xspace}
\newcommand{\commonAppsAll}[0]{20\xspace}

\newcommand{\iraniansCommonsApps}[0]{45\xspace}

\newcommand{\sibappTotalDownloads}[0]{22,989,480\xspace}
\newcommand{\iappsTotalDownloads}[0]{15,253,803\xspace}

\newcommand{\numOfAllPaidStoreA}[0]{20\xspace}

\newcommand{\sibappTotalCracked}[0]{65\xspace}
\newcommand{\iappsTotalCracked}[0]{99\xspace}
\newcommand{\appstarTotalCracked}[0]{18\xspace}

\newcommand{\sibappTotalUpdates}[0]{234\xspace}
\newcommand{\iappsTotalUpdates}[0]{445\xspace}
\newcommand{\appstarTotalUpdates}[0]{206\xspace}
\newcommand{\observedDownloadStoreB}[0]{1,037,935\xspace}
\newcommand{\observedDownloadStoreA}[0]{1,356,210\xspace}
\newcommand{\totalUniqueApps}[0]{1767\xspace}
\newcommand{\totalIran}[0]{510\xspace}
\newcommand{\totalPaid}[0]{308\xspace}
\newcommand{\totalCracked}[0]{159\xspace}
\newcommand{\totalNewAdded}[0]{826\xspace}
\newcommand{\APPUPDATES}[0]{885\xspace}
\newcommand{\OBSEREVEDDOWNLOADS}[0]{2.4M\xspace}
\newcommand{\duration}[0]{5.5\xspace}

\newcommand{\analyzedApps}[0]{568\xspace}
\newcommand{\fileDiffApps}[0]{530\xspace}
\newcommand{\crackToolApps}[0]{360\xspace}
\newcommand{\binaryCompApps}[0]{475\xspace}
\newcommand{\atsCompApps}[0]{531\xspace}
\newcommand{\atsDiffApps}[0]{12\xspace}
\newcommand{\atsWeakerApps}[0]{10\xspace}

\newcommand{\permDiffApps}[0]{3\xspace}

\newcommand{\totalFlows}[0]{52,067\xspace}
\newcommand{\httpsFlows}[0]{51,640\xspace}
\newcommand{\plaintextFlows}[0]{427\xspace}
\newcommand{\plaintextPct}[0]{0.82\%\xspace}
\newcommand{\appsWithFlowsAppStore}[0]{488\xspace}
\newcommand{\appsWithFlowsIApps}[0]{293\xspace}
\newcommand{\appsWithFlowsSibApp}[0]{71\xspace}

\newcommand{\flowPairsCompared}[0]{435\xspace}
\newcommand{\plaintextApps}[0]{77\xspace}

\usepackage{color}
\usepackage{xcolor}
\usepackage{xspace}
\usepackage{url}
\usepackage{subcaption}
\usepackage[font={footnotesize},labelfont=bf]{caption} 
\usepackage{pifont} 
\usepackage{amsmath}

\usepackage[normalem]{ulem} 

\usepackage{enumitem} 
\setlist{nosep} 
\setlist{noitemsep} 
\setlist[itemize]{leftmargin=*} 

\usepackage{booktabs} 

\usepackage{array} 
\newcommand{\PreserveBackslash}[1]{\let\temp=\\#1\let\\=\temp}
\newcolumntype{C}[1]{>{\PreserveBackslash\centering}p{#1}}
\newcolumntype{R}[1]{>{\PreserveBackslash\raggedleft}p{#1}}
\newcolumntype{L}[1]{>{\PreserveBackslash\raggedright}p{#1}}

\usepackage{lipsum}

\makeatletter
\renewcommand\section{\def\@toclevel{1}%
  \@startsection{section}{1}{\z@}%
  {1pt plus 2pt minus 2pt}%
  {1pt plus 1pt minus 2pt}%
  {\ACM@NRadjust\@secfont}}
\let\ACM@origsection\section
\renewcommand\subsection{\def\@toclevel{2}%
  \@startsection{subsection}{2}{\z@}%
  {1pt plus 2pt minus 2pt}%
  {1pt plus 1pt minus 2pt}%
  {\ACM@NRadjust\@subsecfont}}
\let\ACM@origsubsection\subsection
\renewcommand\subsubsection{\def\@toclevel{3}%
  \@startsection{subsubsection}{3}{\z@}%
  {1pt plus 2pt minus 2pt}%
  {1pt plus 1pt minus 2pt}%
  {\ACM@NRadjust{\@subsubsecfont\@adddotafter}}}
\let\ACM@origsubsubsection\subsubsection
\makeatother

\newcommand{\sys}[0]{Havva\xspace}

\newif\ifsubmit
\submitfalse

\ifsubmit
    \newcommand{\amir}[1]{}
    \newcommand{\ah}[1]{}
    
    \newcommand{\mytodo}[1]{}
    \newcommand{\axout}[1]{}
\else
    \newcommand{\axout}[1]{\textcolor{red}{\xout{#1}}}
    \newcommand{\amir}[1]{\textcolor{magenta}{Amir: #1}}
    \newcommand{\ah}[1]{\textcolor{blue}{Amirhosein: #1}}
    
    \newcommand{\mytodo}[1]{\textcolor{red}{TODO: #1}}
    
\fi

\newcommand{\paratitle}[1]{\noindent\textbf{\textit{#1.}}\xspace}

\newcommand{\ie}[0]{\emph{i.e.,}\xspace}
\newcommand{\etal}[0]{\emph{et~al.}\xspace}
\newcommand{\eg}[0]{\emph{e.g.,}\xspace}

\begin{document}
\title{Taking a Bite Out of the Forbidden Fruit:\\
Characterizing Third-Party Iranian iOS App Stores}

\author{Amirhossein Khanlari}
\affiliation{%
  \institution{Stony Brook University}
  \country{New York, USA}
}
\email{akhanlari@cs.stonybrook.edu}

\author{Amir Rahmati}
\affiliation{%
  \institution{Stony Brook University}
  \country{New York, USA}
}
\email{amir@rahmati.com}
\begin{abstract}
Due to U.S. sanctions and strict internet censorship, Iranian iOS users are barred from accessing the Apple App Store and developer services. In response, despite violating Apple’s developer terms, a thriving underground ecosystem of third-party iOS app stores has emerged to serve Iranian users. This paper presents the first comprehensive empirical study of these clandestine app stores. We document how these stores operate, including their distribution mechanisms, user authentication processes, and evasion techniques. By collecting and analyzing more than 1700 iOS application packages and their metadata from three major Iranian third-party app stores, we characterize the ecosystem’s size, structure, and content. Our analysis reveals a significant presence of Iranian-exclusive apps, widespread distribution of cracked apps, unauthorized monetization of paid content, and embedded third-party tracking and piracy libraries. We also uncover a notable overlap among financial, navigational, and social apps that exist solely in this ecosystem, reflecting the unique digital constraints of Iranian users. Finally, we quantify the potential revenue losses for developers due to piracy and document security and privacy risks associated with altered binaries. Our findings highlight how sanctions, censorship, and enforcement gaps have enabled a parallel app distribution ecosystem with complex socio-technical implications.
\end{abstract}
\maketitle

\section{Introduction}

Iranian iOS users and developers face a unique set of challenges stemming from the intersection of international sanctions and strict domestic internet censorship. Since the imposition of U.S. sanctions~\cite{iran_sanctions} and the subsequent tightening of digital restrictions by Iranian authorities~\cite{irancensor}, access to the official Apple App Store and related developer services has been systematically blocked for individuals residing in Iran. This exclusion not only prevents ordinary users from creating Apple Accounts to download, update, or purchase both free and paid apps but also effectively bars Iranian developers from distributing their products on a global platform. As a result, a parallel ecosystem of clandestine third-party iOS app stores has emerged to fill this market gap. These underground platforms fill a critical void but also introduce a host of new risks, including unauthorized app distribution, widespread piracy, and significant security and privacy vulnerabilities for users.
Furthermore, the expansion of unauthorized app stores undermines the safety and reliability of the mobile application landscape for millions of Iranians, exposing them to altered, outdated, or potentially malicious binaries that Apple’s review processes would otherwise filter out. For developers, these stores facilitate the unauthorized monetization of paid content and the widespread distribution of cracked apps, resulting in substantial revenue losses and further discouraging legitimate development efforts in the region.
\par Previous studies have explored mobile application ecosystems, app security, and the effects of sanctions on digital services. Research has focused on app removal dynamics~\cite{removed_ios}, the accuracy of privacy labels~\cite{ios_privacy_labels}, and malware detection~\cite{crios,pios} in both official~\cite{same_app} and alternative~\cite{hey_you} Android markets. In the Iranian context, research on third-party Android app stores, such as Cafebazaar~\cite{mobile_downloads_cafe}, has examined user behavior and app quality. Broader studies~\cite{internet_censorship_in_iran} have investigated the impact of internet censorship on access to information and online services. However, there remains a substantial gap in the literature regarding the operation, content, and risks of clandestine third-party iOS app stores. Specifically, no prior research has systematically documented how these platforms operate, the nature of their app inventories, or the implications for user security and developer revenue on iOS devices.
\par To address these gaps, this paper presents \sys\footnote{\sys is the Persian name for Eve}, the first system to enable the comprehensive empirical study of clandestine iOS app stores. Using \sys, we systematically characterize the operational practices, distribution mechanisms, and evasion techniques employed by three popular clandestine Iranian iOS app stores, and analyze the scale, structure, and content of their app inventories. We choose to focus our work toward the iOS ecosystem, which is often overlooked in favor of Android \cite{ios_overlooked}.
We provide a data-driven analysis that quantifies both the scope of the problem and its broader implications. Furthermore, we directly compare apps obtained from third-party sources with those from official channels, revealing the unique socio-technical dynamics that have emerged in response to sanctions and censorship within the iOS ecosystem.
\par \sys combines several complementary approaches to ensure a fully automated, robust, and thorough analysis. It incorporates custom web crawlers to systematically collect metadata and application packages from three major Iranian third-party iOS app stores, capturing a wide range of information, including app names, versions, categories, user ratings, and download statistics, as well as tracking update frequencies, pricing strategies, and the prevalence of cracked or pirated content. To supplement this, we utilized the iTunes API and custom scripts to retrieve data on corresponding apps from the official App Store, enabling direct cross-platform comparisons. \sys decrypts IPA binaries using a companion jailbroken device, allowing for in-depth static analysis.
This approach enables \sys to examine app characteristics, external libraries, embedded domains, and binary modifications.
\par Our results reveal a robust and resilient parallel app market that is both a response to and a consequence of sanctions and censorship. We document a significant presence of Iranian-exclusive apps. Approximately 510 apps out of \totalUniqueApps analyzed (nearly 29\%)—many of which are essential for the daily life of Iranians but are absent from the official App Store due to policy restrictions. This ecosystem also features the widespread distribution of global apps in cracked or pirated forms, with at least \totalCracked cracked apps and \totalPaid paid apps circulating without authorization, free of charge. Notably, we identify substantial overlaps among apps offered by the stores in critical categories such as finance, navigation, and social networking, leaving Iranian users with no alternative but to rely on these third-party platforms. Our analysis quantifies the potential revenue losses for developers, estimating that piracy and unauthorized distribution in just two stores could exceed \$5.26 million in lost revenue. We also highlight security and privacy risks, with static analysis revealing that 489 apps included unauthorized provisioning profiles and over 180 apps contained piracy libraries such as Cydia Substrate and iGameGod. These findings underscore the complex interplay between geopolitical policy, local innovation, and user behavior and advance our understanding of how digital markets adapt and evolve under conditions of constraint. Ultimately, this work provides a foundation for future research and policy discussions on the unintended consequences of sanctions and censorship in the global digital ecosystem.

Our contributions are summarized as follows:
\begin{itemize}[noitemsep,topsep=0pt]
    \item We develop \sys, the first system to enable a comprehensive empirical study of clandestine iOS app stores.
    \item We study three clandestine Iranian iOS app stores over \duration months, gaining unique insights into their operational practices, app catalog, and user base.
    \item We explore the cracked application catalog in these stores to quantify the potential revenue losses and analyze altered binaries and embedded third-party tracking libraries. 
\end{itemize}
To facilitate future research, our dataset will be available to other researchers upon request, subject to fair use provisions and appropriate data use agreements. For more information, please view the appendix. 

\section{Background}
\subsection{Sanctions}
Since the 1979 revolution and the breakdown of diplomatic ties with the US, Iran has faced various sanctions~\cite{iran_sanctions,sanctions_timeline}. Over time, these sanctions have progressively broadened through multiple executive orders. The most impactful of which were introduced after the end of the Iran-Iraq War in the 1990s.  The next wave of sanctions was initiated after the events of September 2001~\cite{sanctions_timeline}, resulting in major technology companies such as Yahoo and Microsoft refraining from providing certain public and free services to Iranian users~\cite{zoomit}.
\par On March 8, 2010, the U.S. Treasury's Office of Foreign Assets Control (OFAC) published a list of sanctions against Iran, reiterating that:
 \begin{quoting}[leftmargin=.5cm, rightmargin=.5cm]
    ``Section 560.204 of the Iranian Transactions Regulations (ITR) provides that the exportation, reexportation, sale, or supply, directly or indirectly, from the United States or by a U.S. person, wherever located, of any goods, technology, or services to Iran or the Government of Iran is prohibited."~\cite{ofac_sanctions}
\end{quoting}

\par As a consequence of these sanctions, certain technological services are restricted in Iran. For example, computers and mobile devices attempting to download from the iTunes Store encounter \texttt{error 1009}, indicating that the country from which they are downloading is on the restricted list~\cite{err1009}. Access to certain Google services, such as Google App Engine and websites that use it (\eg Khan Academy) was blocked for Iranian users~\cite{zoomit}. Similarly, two-factor authentication is not available on most online services because they lack support for Iranian phone numbers. To limit the harm to regular users, OFAC issued General License D-2 in September 2022, permitting export of specific services, software, and hardware incident to personal communications, such as instant messaging, chat, and email to Iran:
 \begin{quoting}[leftmargin=.5cm, rightmargin=.5cm]
"Section 560.540(a)(2): The exportation or reexportation [...] of services, software, and hardware incident to the exchange of personal communications over the Internet [...] is authorized.”~\cite{license_d2}    
\end{quoting}
 Despite this exception, access to many services remains limited due to the inconsistent implementation of this exception by technology companies.
\par On the economic front, Iranian entities cannot access the SWIFT international payment network, severely complicating cross-border payment processing~\cite{swift_remove}. Consequently, payment systems such as Visa, Mastercard, and PayPal are unavailable in Iran. This restriction significantly hinders Iranian citizens from making online purchases, subscribing to services, or participating in global e-commerce. Iranians frequently circumvent this restriction by using gift cards or opening bank accounts in foreign countries that offer Visa or Mastercard services.

\subsection{Censorship}
Iran enforces one of the strictest internet censorship regimes in the world~\cite{irancensor}. About 70\% of Internet traffic is blocked in Iran \cite{70percent_iran_traffic}. 49/100 of the world's most popular websites are inaccessible in Iran, second only to China~\cite{censor_percentage}. Popular services, including YouTube, Facebook, Twitter, Instagram, and Telegram, are blocked within Iran~\cite{iran_telegram_instagram,internet_censorship_in_iran,freedom_on_the_internet}. Iranians also experience near-complete Internet blackouts during times of political turmoil and unrest. Most recently, on 8 January 2026, during the 2026 nationwide protest~\cite{iran_jan_protest}, Iranian authorities cut off most internet and mobile connectivity nationwide~\cite {internet_blackout_2026}, and multiple independent monitors such as Cloudflare reported that Iran’s internet traffic dropped to near zero~\cite {internet_blackout_2026_cloudflare,internet_blackout_2026_accessnow}. The Iranian regime partially restored this blackout on January 23rd after quashing these protests~\cite{internet_blackout_restored}, but enforced another nationawide blackout after the start of the US-Iran war on February 28th~\cite{us_iran_war} which has continued to this day. Other major shutdown episodes include 2025 Iran-Israel war~\cite{war_bloackout}, the 2022~\cite{2022blackout} and 2019 protests~\cite{2019blackout}, and the 2013 Presidential elections~\cite{irancensor}. More than 80\% of Iranians use a VPN to access these websites and services \cite{80percent_vpn}. Access to most major free VPN services and censorship circumvention tools such as TOR is blocked, and demand for such services is constant and ever-increasing~\cite{vpn_surges}. Many ISPs have blocked access to the App Store and Google Play \cite{appstore_gp_censor}, making it more difficult for Iranians to obtain VPN apps, among others.

\subsection{The Apple App Store}
The Apple App Store is the primary digital marketplace for Apple devices, and a platform for discovering, downloading, and purchasing apps within the Apple ecosystem, including iOS, iPadOS, watchOS, and tvOS. 
According to Apple's 2024 transparency report, the App Store hosts over 1.9 million apps and has facilitated more than 800 million downloads weekly~\cite{apple_transparency_report}.
To purchase and download apps from the App Store, users must first create an Apple Account (\ie an Apple ID)~\cite{apple_account,apple_id}. Developers wishing to offer apps on the App Store must enroll in the Apple Developer Program~\cite{developer_benefits}.

Apple’s terms and conditions governing the creation and use of Apple Accounts describe its compliance process with U.S. sanctions. According to the Apple Media Services Terms and Conditions \cite{apple_legal}, individuals residing in countries or regions subject to comprehensive U.S. sanctions—including Iran, Syria, North Korea, Sudan, and Cuba—are prohibited from creating an Apple Account or accessing certain App Store services. The relevant language states:
 \begin{quoting}[leftmargin=.5cm, rightmargin=.5cm]
 “You may not use or otherwise export or re-export the Services except as authorized by United States law and the laws of the jurisdiction in which the Services were obtained. In particular, but without limitation, the Services may not be exported or re-exported to any country to which the United States has embargoed goods or to anyone on the U.S. Treasury Department’s list of Specially Designated Nationals or the U.S. Department of Commerce Denied Person’s List or Entity List.”
 \end{quoting}

Although these changes eased parts of the sanctions framework, they did not by themselves create direct App Store access for users located in Iran.
\par Apple utilizes digital rights management (DRM) systems, such as FairPlay~\cite{fairplay}, that require a valid, region-compliant Apple Account to sign and authorize downloaded apps. Although U.S. sanctions have been eased to allow the export of Apple hardware and software to Iran, Apple’s own account creation policies and technical requirements persist in preventing Iranian residents from officially creating an Apple Account by restricting the selection of Iran as country during account registration and also verifying the user's IP address and displaying \texttt{error 1009} if the IP is identified as originating from Iran. As of today, there is currently no official or legal means for residents of Iran to create an Apple Account. This prohibition affects both developers and ordinary users, effectively excluding Iranian app developers from distributing apps on the platform and hindering end users from downloading both free and paid apps. 
While some users try to bypass these restrictions through technical means, \eg using VPNs or registering Apple Accounts with foreign addresses, these methods do not offer official access and violate Apple’s terms of service.

\begin{figure}[tbh!]
    \centering
    \includegraphics[width=\linewidth]{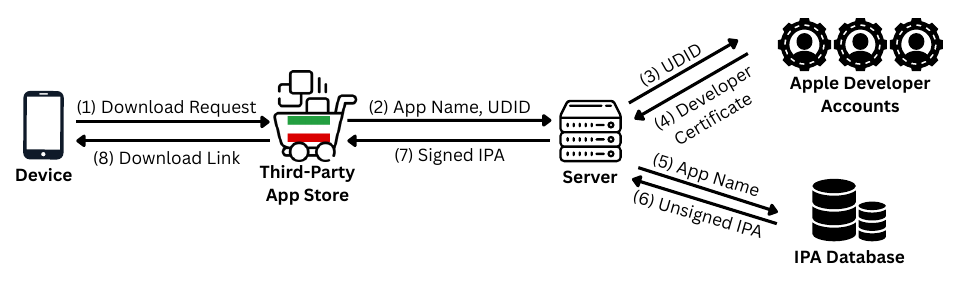}
    \caption{App download flow for third-party stores. A user’s download request in the Ad-hoc distribution model triggers the app store to fetch the developer certificate associated with the developer account that contains the device UDID and the unsigned IPA, and to pass it to the user after signing. In the Enterprise distribution model, all apps are signed with the enterprise certificate, skipping steps 3-6.} \vspace{-1.5em}
    \label{fig:app-dl-flow}
\end{figure}

\begin{figure}[tbh!]
    \centering
    \includegraphics[width=.9\linewidth]{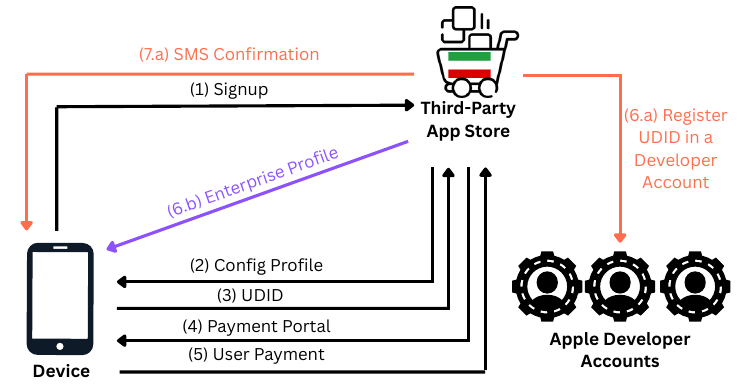}
    \caption{Sign up flow for third-party stores. Black, orange, and purple lines indicate shared, ad hoc-only, and enterprise-only steps, respectively. The device initiates signup and exchanges configuration, payment, and UDID data with the app store, which then registers the UDID with Apple developer accounts or provides an enterprise profile, completing the process with SMS confirmation.}
    \label{fig:signup-flow}
\end{figure}

\subsection{Apple Developer Program}
Installing apps without using the App Store involves sideloading. Sideloading refers to the process of installing apps on devices from sources outside the official App Store~\cite{sideloading}. Sideloading can be done through multiple distribution channels inside the Apple developer ecosystem.
The Apple Developer Program is a subscription-based platform that provides users with access to developer tools, including Xcode, Swift, multiple APIs, and resources necessary for creating, testing, and distributing apps across the Apple ecosystem~\cite{developer_benefits}. This program requires an annual subscription of \$99 for both individuals and organizations~\cite{dev_price} and provides developers with six paths to distribute their apps~\cite{distribute_beta}:\\
\paratitle{Developer Distribution} which is designed to test apps at a very small scale on devices owned by developer teams during the development phase. Sideloading using developer distribution can be performed by anyone, even without a paid developer account. However, it is restricted to three apps per device per week, with a seven-day expiration period for apps installed under a free developer account~\cite{distribute_beta}. Using this method, users can sign and install any apps using tools such as AltStore~\cite{altstore} and Sideloadly~\cite{sideloadly}.

\paratitle{Ad-Hoc Distribution} is similarly used for testing, albeit at a larger scale. It allows developers to distribute a pre-release version of their apps to up to 100 iPhones and 100 iPads per year (per developer account)~\cite{devices_overview}. To distribute an app to a target device, the device's UDID (Unique Device Identifier) must be registered inside the Apple Developer Account~\cite{distributing_resigtered}. Apps must be signed using a provisioning profile generated through the Apple Developer portal to facilitate installation on target devices. This signed bundle is primarily installed on registered devices through Apple's wireless app installation method~\cite{wireless_distribution}.

\paratitle {Enterprise Distribution} This method requires a specific subscription, the Apple Enterprise Program, which costs \$299 per year \cite{dev_price}. Only legal entities such as companies, organizations, and institutions are eligible to enroll in this program. A DUNS (Data Universal Numbering System) Number is required for getting verified by Apple \cite{enterprise_program}. 
Once an enterprise account is obtained, the developer can sign apps using an enterprise certificate. Apps signed by an enterprise certificate can be installed on any iPhone and iPad by manually trusting the organization profile without requiring registration of their UDID in the Apple Developer Account \cite{enterprise_distribution}.

\paratitle{TestFlight} allows for testing app's beta version with limited Over-the-Air (OTA) users. TestFlight allows developers to gather feedback and insights from testers before releasing the app on the App Store. It permits developers to evaluate their applications with up to 100 internal users within the development team and up to 10,000 external testers per app. External testers are provided with an invitation link by the developer via email or public links. Each build intended for external testers must be reviewed and approved by Apple before distribution. Each build remains valid for 90 days \cite{test_flight}.

\paratitle{App Store Distribution} Requires developers to submit their app through App Store Connect \cite{appstore_connect}, where it must pass the App Review Process~\cite{app_review} and checked for compliance with the App Store Guidelines~\cite{review_guideline}. Once approved, the developer can publish their apps on the App Store.

\paratitle{European Union Distribution} This method was introduced in March 2024 with iOS 17.4~\cite{european_union}, after a lawsuit against Apple and Meta in Europe \cite{european_lawsuit}. It allows apps to be distributed outside Apple’s official App Store in Third-Party stores within the EU. To access these third-party stores, users must have an Apple Account in the European Union, and their devices should be physically located in European countries \cite{installing_european}.

\section{How Clandestine App Stores Work}
Clandestine third-party app stores typically use either the ad-hoc or enterprise distribution path to reach their user base. In this section, we discuss the process of signing up for and using these app stores under both distribution paths.

\paratitle{Sign up Process}
The registration procedure in Iranian third-party stores is illustrated in Figure~\ref{fig:signup-flow}. Depending on the distribution method employed by the store, there are two possible flows: enterprise distribution or ad-hoc distribution.

\par These stores are designed to provide service exclusively to Iranian individuals. The initial registration requires possession of an Iranian phone number, as a verification code is sent for confirmation. Upon entering the user's phone number, the user is prompted to download a profile designed to extract the device UDID. This UDID is utilized to determine whether the user has previously subscribed. If not, the user is prompted to purchase a subscription, which necessitates an Iranian payment method and an Iranian IP address. Following a successful payment, depending on the distribution path, the flow will diverge. For the ad-hoc distribution path, a waiting period of up to 72 hours is triggered, allowing the app store owner to register the device's UDID in one of the store's Apple developer accounts~\cite{udid_time}. Following the period, users receive a text message and can download apps from their platform. 
When the store uses the enterprise distribution path, an enterprise profile will be delivered to the device instead. This allows the installation of any app with a pre-signed IPA on all devices.

\begin{figure}[tb]
    \centering
    \includegraphics[width=\linewidth]{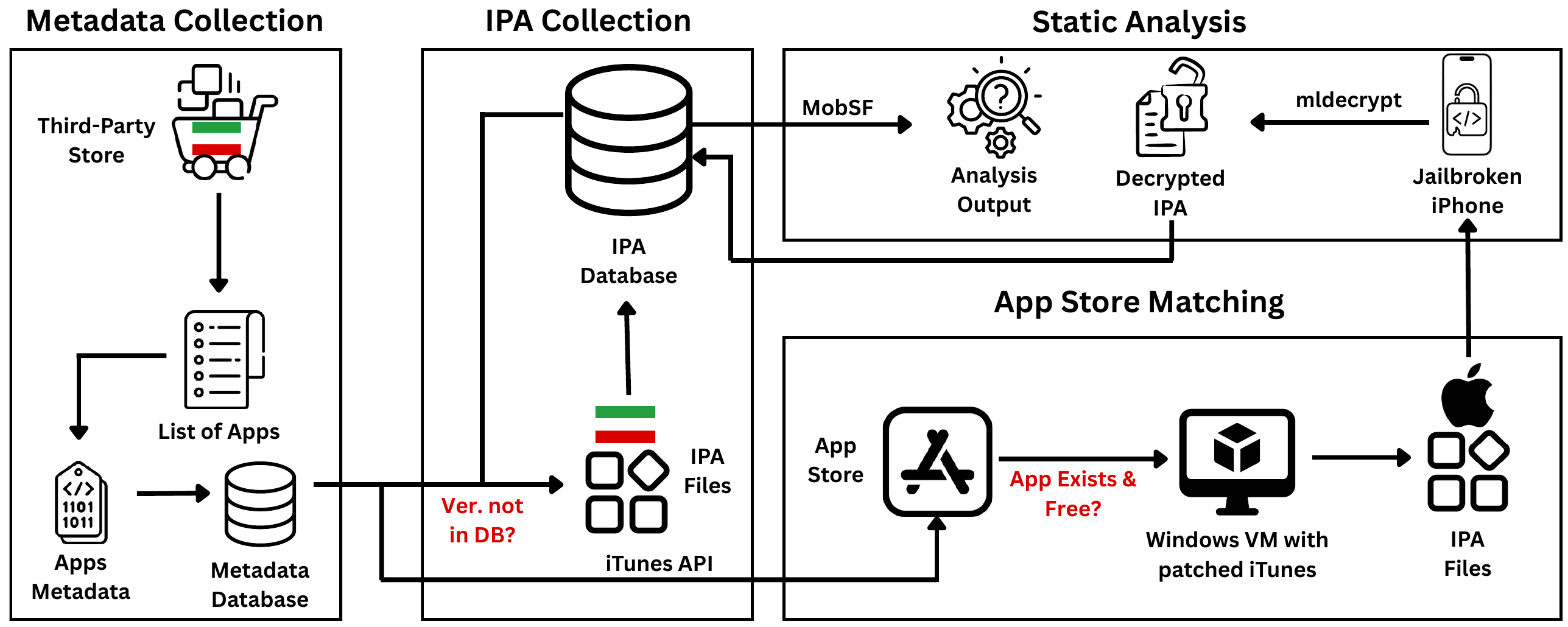}
    \caption{\sys collects app metadata and IPA files from third-party stores and correlates them with apps in the App Store. It then checks for new versions and stores them in a database. IPA files collected from the App Store are decrypted and used to perform comparative analysis with IPA files collected from third-party stores.}
    \label{fig:sys-flow}
\end{figure}

\paratitle{App Download}
The app download process is demonstrated in Figure~\ref{fig:app-dl-flow}. When a subscribed user attempts to download or update an app from a store that uses Ad-Hoc distribution, the developer account associated with the UDID will be identified first. Third-party stores retain all decrypted IPA files for their apps within their database. Subsequently, the IPA file for that app will be signed using the ad-hoc provisioning certificate for the developer account to which the device UDID was assigned during the signup step. The duration of this process depends on the size of the IPA file. Upon completion, the user will receive a notification prompting them to accept or decline an OTA installation from that source. Once accepted, the installation will commence on the user's device. 

When a store uses enterprise distribution, there’s no need to match the user’s UDID with a specific developer account since all apps are signed using the same certificate. Users only need to download an enterprise profile to install apps associated with that account, skipping steps 3-6 in the figure.
\par Both these approaches explicitly violate Apple regulations regarding ad-hoc distribution, as outlined in section 7.3 (Ad Hoc Distribution) and 7.6 (No Other Distribution Authorized) \cite{developer_agreement}:
``You agree not to distribute Your Application for iOS, iPadOS, tvOS, visionOS, or watchOS to third parties via other distribution methods or to enable or permit others to do so.''
\par Apple frequently detects such fraudulent accounts and revokes their certifications and privileges~\cite{revokation}. When this happens, users lose access to all downloaded apps from the affected platform. To continue operations, these stores purchase new developer accounts, update their signing mechanisms with new certificates, and require users to re-download the previously revoked apps.

\section{\sys's Design}

The high-level architecture of \sys is presented in Figure~\ref{fig:sys-flow}. \sys is comprised of four modules: (1)~Metadata Collection, (2)~IPA Collection, (3)~App Store Matching, and (4)~Static Analysis. We discuss each of these components in detail below.

\subsection{Metadata Collection}
\sys utilizes custom-developed web crawlers designed to extract relevant information from both third-party and official stores. These web crawlers are engineered to crawl store webpages and retrieve app listings. These lists are transmitted to another crawler, which is designed to extract app metadata, such as Bundle ID (a unique identifier assigned to each app on the App Store), Version, Size, Download Count, and other relevant details, from individual app pages via the stores' APIs. The extracted metadata is systematically incorporated into dedicated databases for each store, along with the corresponding date of data retrieval.

\subsection{IPA Collection}
iOS apps are distributed as IPA (iOS Package Archive) files. For each \texttt{[App, Ver]} record in the metadata database, \sys first verifies whether the exact version of the app exists within the IPA database. If it does, it skips this record; if not, \sys acquires the corresponding IPA file. These stores do not offer public APIs for downloading IPA files. Consequently, we establish a Man-in-the-Middle proxy on our device to intercept and capture their private APIs.
Utilizing these private APIs, \sys is able to download IPA files from third-party stores.

\subsection{App Store Matching}
For each \texttt{[App,Ver]} pair added to the database, \sys also looks for the matching \texttt{App} in Apple App Store using the App's unique Bundle ID and iTunes API~\cite{itunes_api}. If the app is found, \sys collects the metadata and (when the app is free) the IPA file of the app hosted in the Apple App Store. In many cases, however, the \texttt{[Ver]} available in the Apple App Store may differ from that offered in a third-party store (\ie the third-party store may offer an outdated version of the app).  To extract the correct app version from Apple App Store, we employed the NyaMisty script~\cite{pached_itunes} along with a Windows Server 2022 Virtual Machine and a modified version of iTunes v12.6.5.3, which allows \sys to download older versions of these apps. This approach ensures that we can retrieve the correct, unaltered version of the app, which is otherwise inaccessible via the standard iTunes API.

\subsection{Static and Dynamic Analysis}
An IPA bundle can be extracted by renaming its extension from .ipa to .zip. This bundle comprises resources, executables, and libraries required to run an iOS app. The primary executable is a Mach-O file encrypted with FairPlay DRM, thereby preventing it from being executed on unauthorized devices. Third-party stores must sign a decrypted version of this executable to facilitate its signing and sideloading onto their users' devices. \sys similarly requires access to decrypted apps to compare them to their third-party store counterparts. The process of decrypting IPA files necessitates the use of a jailbroken device. For this purpose, we used a jailbroken iPhone 8. The installation of IPAs on the device is accomplished via ideviceinstaller~\cite{ideviceinstaller}, followed by the use of mldecrypt~\cite{mldecrypt} for the decryption of binaries.

\par We used MobSF tools~\cite{mobsf} to perform static analysis on apps obtained from the App Store and third-party stores. Subsequently, we conducted a comparative study of results obtained from decrypted apps in the App Store and from third-party stores. Our analysis covers essential app data, including size, version, bundle identifier, permissions, file listings, external libraries, domains, URLs, binary analysis, traffic flow analysis, and Mach-O analysis.
\subsection{\sys App Coverage}
\sys successfully collected metadata information from all apps present in third-party stores, as well as all matching metadata from the official App Store. However, during the process of gathering the package files of these apps, \sys excluded \totalPaid paid apps from the official App Store due to the associated cost. These omitted applications were excluded from our binary analysis as the original packages were unavailable for comparison. Furthermore, because our jailbroken iPhone was running iOS 16.7.12, we were limited to decrypting apps that did not require iOS 17 or later.
Consequently, 10 apps were excluded from our static analysis.
\section{Third-Party App Store Analysis}
In this section, we provide a comprehensive analysis of our dataset of three third-party app stores. We selected these stores based on their popularity, app diversity, and the presence of cracked apps~\cite{iranian_stores_ranking}. In our app breakdown, we define \textit{Iranian apps} as applications that are not available in the official app store and are exclusively available in third-party stores. In our manual inspection, we observed that, in addition to Iranian apps, this list includes a few apps that have been removed from the app store but remain available in third-party stores (\eg Assassin's Creed Identity~\cite{assasin}, Flappy Bird~\cite{flappy_bird}). In contrast, we define \textit{global apps} as applications that are also available in the official Apple App Store. We also define \textit{cracked apps} as applications that require a paid subscription to access all functionalities but have been modified to bypass these requirements.
\par Between April 14th 2025 and October 1st 2025, we collected \totalUniqueApps apps including \totalIran Iranian apps, \totalPaid paid apps, and \totalCracked cracked apps. In this period, we observed \totalNewAdded new apps, \APPUPDATES app updates, and over \OBSEREVEDDOWNLOADS new downloads (Only two of the three stores provide download statistics). Table~\ref{tab:overview} provides a high-level view of our dataset. 

\begin{table}[tb]
\footnotesize
\centering
\caption{General statistics of three major Iranian third-party app stores. ``Observed'' rows describe the number of occurrences during the measurement period. \appstar does not provide download numbers.}
\begin{tabular}{@{}llll@{}}
\toprule
                      & \textbf{\sibapp} & \textbf{\iapps} & \textbf{\appstar} \\ \midrule
Membership Cost (3 Months)       &2,970K IRR         &5,390K IRR          &2,990K IRR          \\ \midrule
Total Apps            & \sibappTotal        &\iappsTotal         & \appstarTotal        \\ \midrule
 Free Global Apps            & 236       &800         & 49        \\ \midrule
 Paid Global Apps            & 68       &195         & 74       \\ \midrule
Iranian Apps            & 254        &328         & 98        \\ \midrule
Cracked Apps    &\sibappTotalCracked         &\iappsTotalCracked         &\appstarTotalCracked          \\ \midrule
Observed Update Events &\sibappTotalUpdates       &\iappsTotalUpdates       &\appstarTotalUpdates        \\ \midrule
Observed Downloads & \observedDownloadStoreA       & \observedDownloadStoreB       & N/A       \\ \midrule
Total Downloads       &\sibappTotalDownloads         &\iappsTotalDownloads         & N/A         \\ \bottomrule
\end{tabular}
\label{tab:overview}

\end{table}

\subsection{App Overlap Between Stores}

\begin{figure}[tb]
    \centering
    \label{fig:venn}
    \includegraphics[width=.6\linewidth]{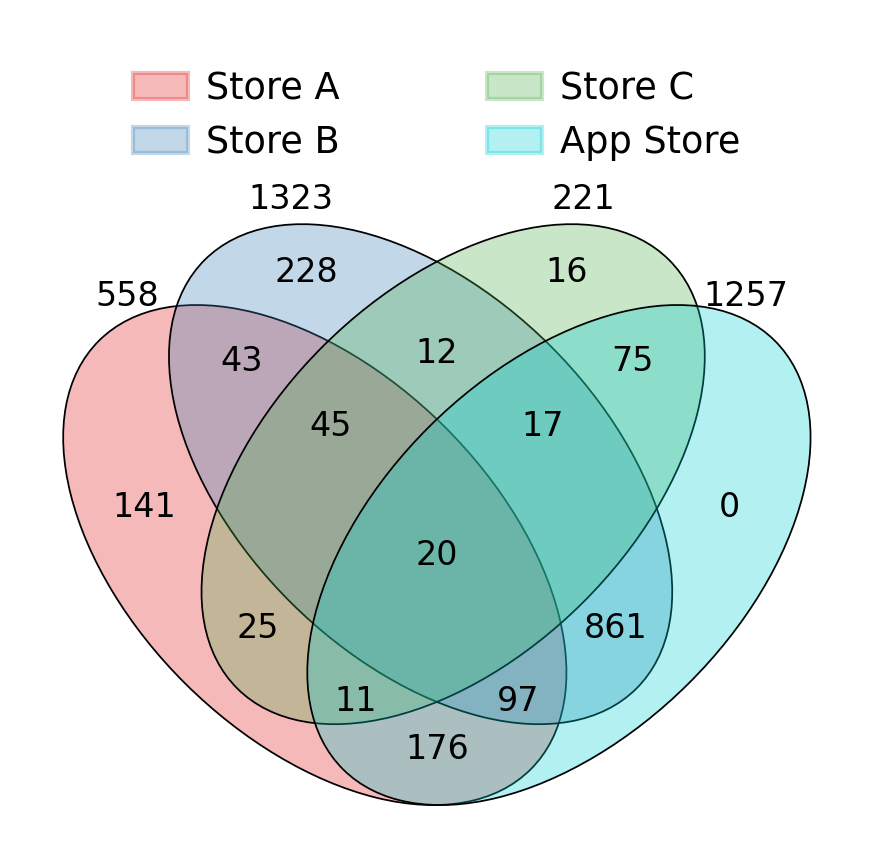}
    \caption{Venn diagram illustrating the numbers of
applications shared among all third-party stores and the official App Store. \vspace{-1em} }
\end{figure}

\begin{figure*}[tbh]
    \centering
    \includegraphics[width=.9\linewidth]{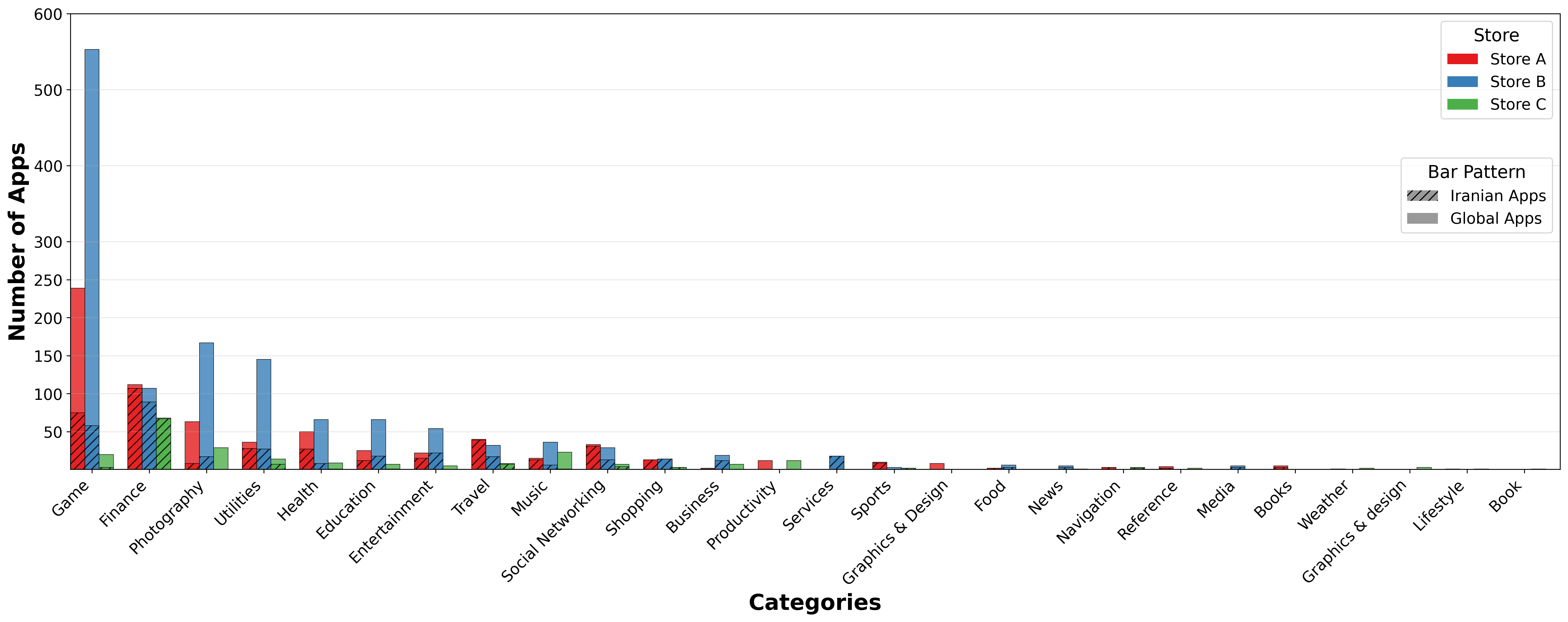}
    \caption{Distribution of apps across various categories, divided by Iranian and global apps \vspace{-2em}}
    \label{fig:app-category}
\end{figure*}

Our findings reveal patterns of pronounced fragmentation and selective overlap within the Iranian app ecosystem. A substantial number of apps are exclusive to the Iranian stores and absent from the Apple App Store. Specifically, \sibapp, \iapps, and \appstar host \sibappTotal, \iappsTotal, and \appstarTotal unique apps, respectively. Many of these apps, not found in the App Store, can be characterized as distinctly Iranian, reflecting local development priorities and user needs. For instance,~\sibappIranianPercentage of \sibapp apps are not available in the App Store. Analysis of intersections further clarifies the distribution of apps. In particular, only \commonAppsAll apps are present across all four platforms, indicating that very few apps achieve universal distribution. The predominant category within these common apps is Photography. Ten photography apps, such as ProCamera and PicsArt, are available across all stores. Most of these apps are paid apps on the App Store and require subscriptions, which are subsequently cracked in Iranian stores. Duolingo and Mondly are two popular free language learning apps that are also cracked in these stores to bypass paid subscriptions. Additionally, Spotify is included on this list, with a cracked subscription available from third-party stores. It can be concluded that the availability of cracked subscriptions significantly contributes to the popularity and ubiquity of these apps across all stores.
In the examination of common apps across Iranian app stores, we identified \iraniansCommonsApps apps. Finance apps constitute 68\% of these apps. The list includes apps for many prominent Iranian banks such as Melli, Saderat, and Saman. The existence of these apps can be attributed to the financial sanctions imposed on Iran and the absence of Iranian apps from official app stores. Iranians are only able to access their local banking apps via third-party stores. Neshan and Balad are the two primary navigation apps available in Iran that are included in this list. Major navigation services such as Waze are restricted in Iran due to censorship, and Google Maps and Apple Maps do not provide effective coverage within the country.

\subsection{Sanctions Evasion via App Morphing}
\begin{figure}[tbh]
    \centering
    \begin{subfigure}[t]{0.45\linewidth}
        \centering
        \includegraphics[height=3in]{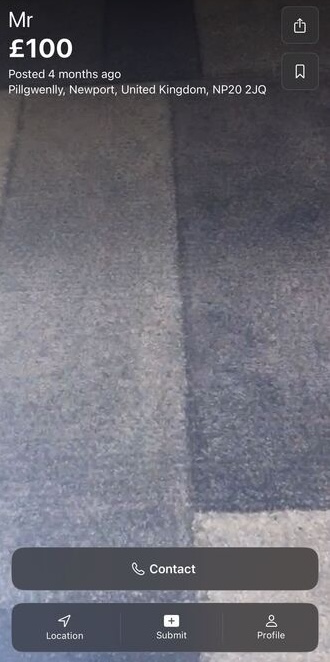}
        \caption{Before detecting Iranian IP}
    \end{subfigure}
    ~ 
    \begin{subfigure}[t]{0.45\linewidth}
        \centering
        \includegraphics[height=3in]{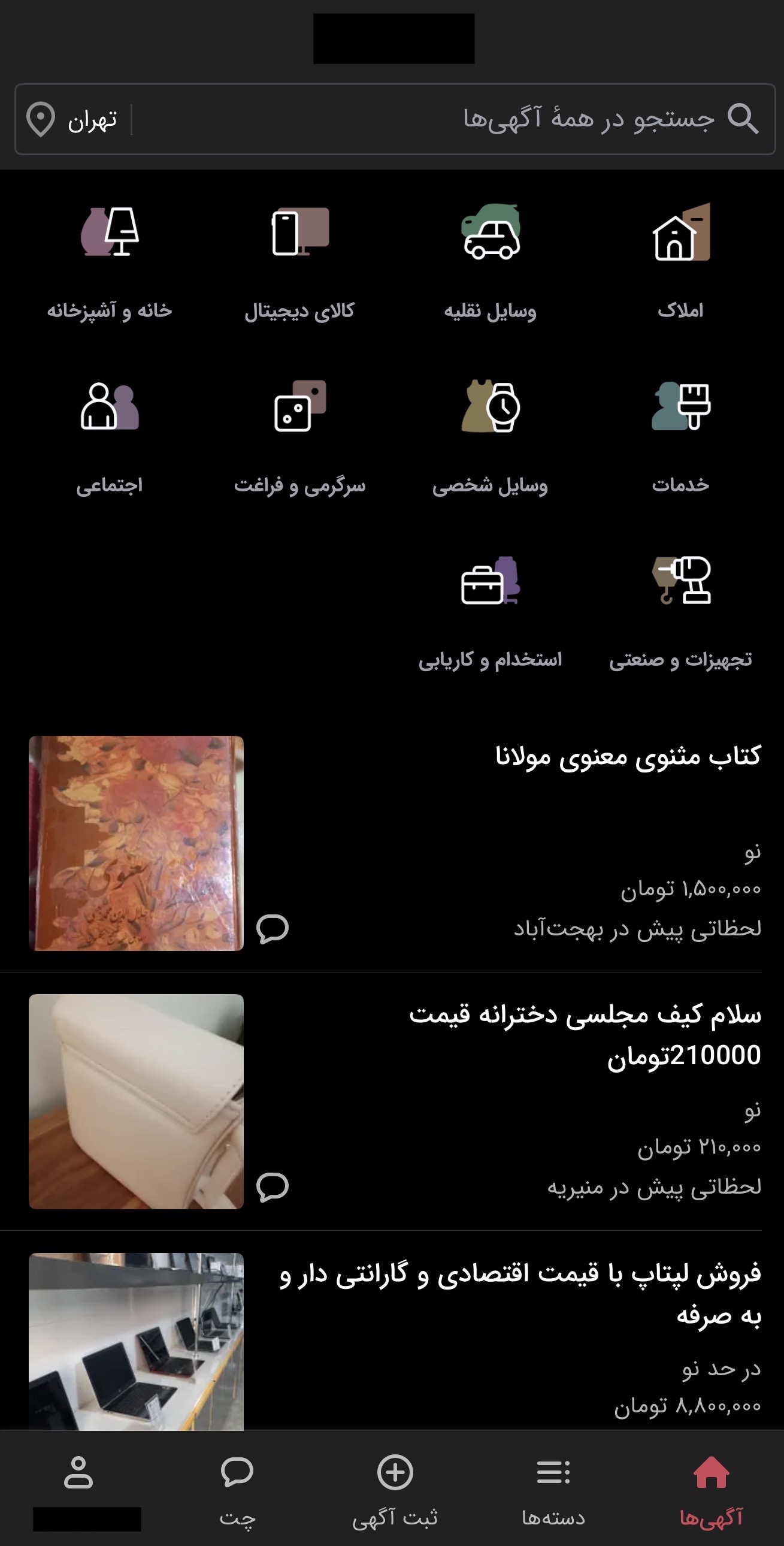}
        \caption{After detecting Iranian IP}
    \end{subfigure}
    \caption{Example of a morphed Iranian application in the Apple App Store. The app appears as a TikTok replicate but morphs into an Iranian e-commerce platform upon detecting Iranian IP}
    \label{fig:morphapp}
\end{figure}

In examining common applications across Iranian stores and the Apple App Store, we discovered 3 unique Iranian apps that were also unexpectedly available on the App Store. Upon manual inspection, we discovered that these apps were shell apps, where they appeared as benign apps (\eg  basic utilities or games) when opened normally, but morphed into the intended Iranian-specific apps when the device presented an IP  located in Iran. 
Figure~\ref{fig:morphapp} presents an example of such an app. When opened outside of Iran, it looks like a simple video-sharing app similar to TikTok (Figure~\ref{fig:morphapp}a), but it morphs into a popular Iranian classified ads platform when accessed from inside Iran (Figure~\ref{fig:morphapp}).
The second app launches as a web browser, but after detecting an Iranian IP address, transformed into a ebook-reading app.
The third app is a navigation app. It launches as a Google Maps replica, but after detecting an Iranian IP, it transforms into a distinct Iranian navigation app.
While app cloaking is typically used by gambling and scam apps to evade store rules globally, its use by prominent Iranian tech companies shows the extreme measures they take to remain present on the official iOS App Store.

\subsection{App Categories}
Figure~\ref{fig:app-category} presents the breakdown of apps in each category.
Our results reveal that Iranian apps predominantly populate categories such as Finance, Social Networking, and Shopping. Finance apps primarily consist of Iranian banking apps, with a limited number of global financial apps. These international financial applications are exclusively for trading and analyzing cryptocurrencies, such as KuCoin and TradingView. Cryptocurrency remains the sole international payment system accessible to Iranians due to sanctions. Popular social networks such as Instagram, Facebook, and Telegram are banned within Iran; consequently, local social networking platforms like Rubika and Eitaa are available in app stores. Iranians lack access to global e-commerce platforms such as Amazon and eBay; instead, they use local online platforms, including Digikala and BaniMode. These trends align with our expectations: globally known commercial networks are unusable to Iranians due to sanctions, while Iranian-specific commercial frameworks are absent from the App Store for the same reason.
\par Categories such as Games, Photography, and Education are dominated by global apps. Games constitute the most popular category in \sibapp and \iapps, comprising the majority of apps, whereas most \appstar apps are finance-related. The Photography category primarily consists of cracked apps across all stores, and \iapps is the leading store in terms of the number of such apps.

\subsection{New Apps}
\begin{figure}[tb]
    \centering
    \includegraphics[width=1\linewidth]{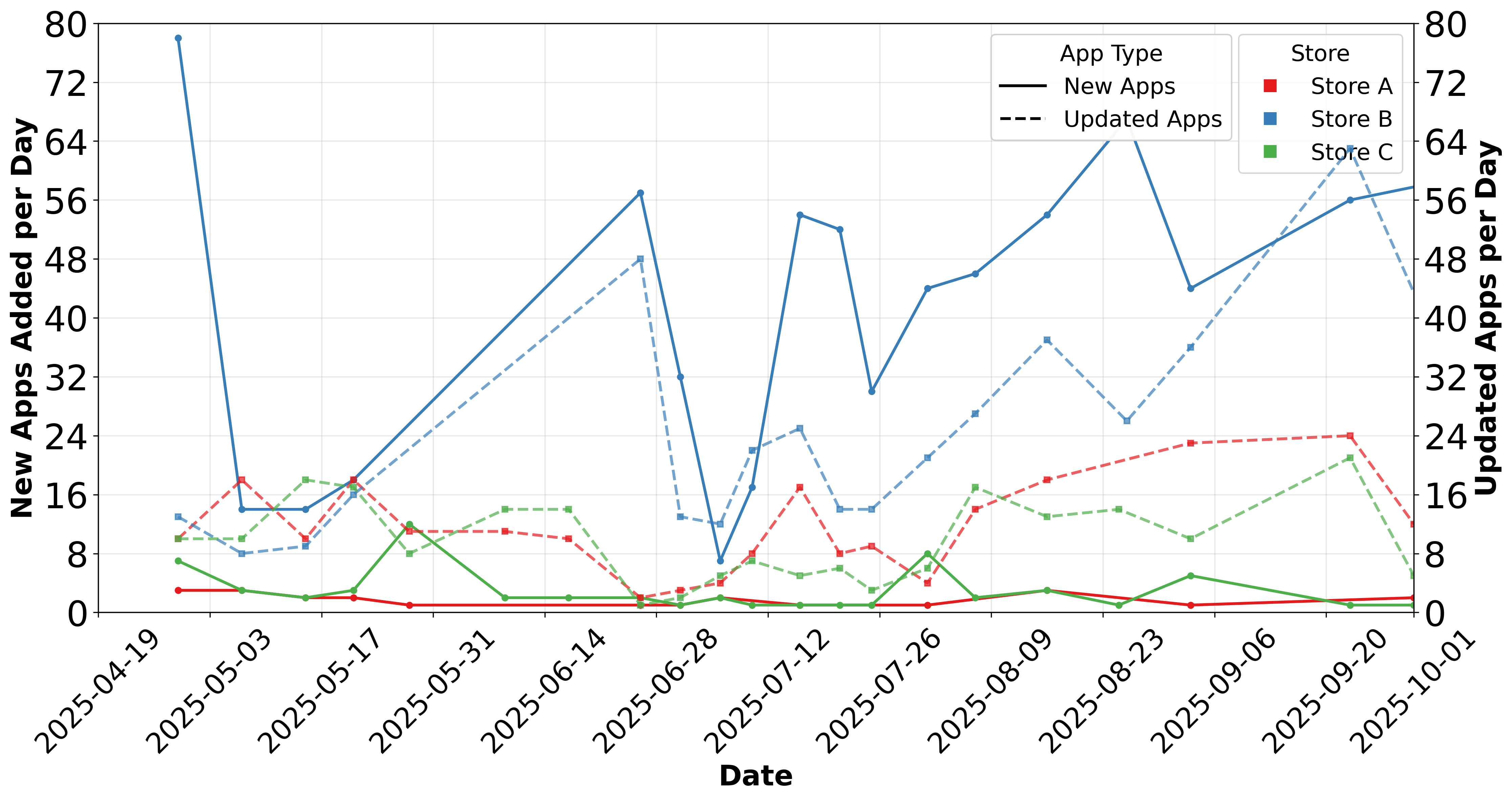}
    \caption{Newly added and updated apps over time.
}
    \label{fig:enter-label}
\end{figure}
During our \duration months data-collection period, we observed \totalNewAdded new apps being added to the three app stores.  Examining these apps across three stores revealed distinct operational strategies and market priorities. \iapps consistently led in app onboarding, with notable peaks of eighty new apps in the week of May 6, 2025, and seventy-two new apps on July 1, 2025, despite not being the most popular store in Iran. This indicates competition among third-party apps and the effort of \sibapp to surpass \iapps in popularity. \appstar exhibited a more moderate and sporadic pattern, maintaining a steady rate of eight new apps per observation early in the period and experiencing a spike to fifteen new apps on the week of May 20, 2025. In contrast, \sibapp, being the most popular store, maintained a consistently low rate of new app additions, with fewer than eight new apps per week throughout the period. This consistent approach indicates a focus on quality and compliance. Furthermore, as illustrated in Fig. 6, \sibapp has the highest number of app updates in most of our observations.
\subsection{App Updates}

\begin{table*}[tb]
\centering
\caption{Top 10 most frequently updated apps across all Iranian stores}
\label{tab:most_updated_apps_combined}
\scriptsize
\resizebox{\textwidth}{!}{
\begin{tabular}{c|ccc|ccc|ccc}

\textbf{Rank} & \multicolumn{3}{c|}{\textbf{\sibapp}} & \multicolumn{3}{c|}{\textbf{\iapps}} & \multicolumn{3}{c}{\textbf{\appstar}} \\
\cline{2-4} \cline{5-7} \cline{8-10}
& \textbf{App} & \textbf{Updates} & \textbf{Description} & \textbf{App} & \textbf{Updates} & \textbf{Description} & \textbf{App} & \textbf{Updates} & \textbf{Description} \\
\hline
1 & Divar & 11 & Craigslist Alternative & Divar & 10 & Craigslist Alternative & Divar & 12 & Craigslist Alternative \\
2 & Shomar & 10 & Banking App & Mencherz & 7 & Persian Game & Bankette & 8 & Banking App \\
3 & Excoino & 9 & Crypto Store & Excoino & 6 & Crypto Store & n-Track Studio Pro & 8 & Acoustic Analysis App \\
4 & Bankette & 7 & Banking App & StepsApp++ & 6 & Fitness App & Audiotools & 8 & Acoustic Analysis App \\
5 & SetareYek & 7 & Banking App & Hamrah Card & 6 & Banking App & Cloud Player Pro & 7 & Video Player \\
6 & AbanTether & 7 & Crypto Store & Boom Beach & 6 & Game & Genius Scan & 7 & Scanner App \\
7 & Nobitex Exchange & 6 & Crypto Store & Home Workout ++ & 5 & Fitness App & FLStudio Mobile HD & 6 & Photography App \\
8 & Hamrah Card  & 6 & Banking App & Gamebox & 5 & Persian Game & Refah Bank & 6 & Banking App\\
9 & Mencherz  & 6 & Persian Game & Minecraft  & 5 & Game & Nobitex & 5 & Crypto Store \\
10 & Apetit Fit & 6 & Persian Fitness App & Homescapes ++ & 5 & Game & Audio Master Pro & 5 & Audio Producing App
\label{tab:most_updated}
\end{tabular}
}
\end{table*}

\paratitle{Most Frequently Updated Apps}
The top 10 most frequently updated apps across three Iranian stores are presented in Table~\ref{tab:most_updated}. The most frequently updated apps in \sibapp are predominantly Iranian, with Divar and Shomar updated 11 and 10 times, respectively. \iapps and \appstar include several global apps in their listings. The most frequently updated global app in \iapps is Mencherz, a popular Persian card game. Among the ten apps in \iapps, only four are Iranian; the remaining are paid or cracked apps. The top 10 frequently updated apps in \appstar largely consist of apps priced between \$20 and \$49.99. Divar is the only app that appears among the most updated apps across all stores.

\begin{figure}[tb]
    \centering
    \includegraphics[width=.9\linewidth]{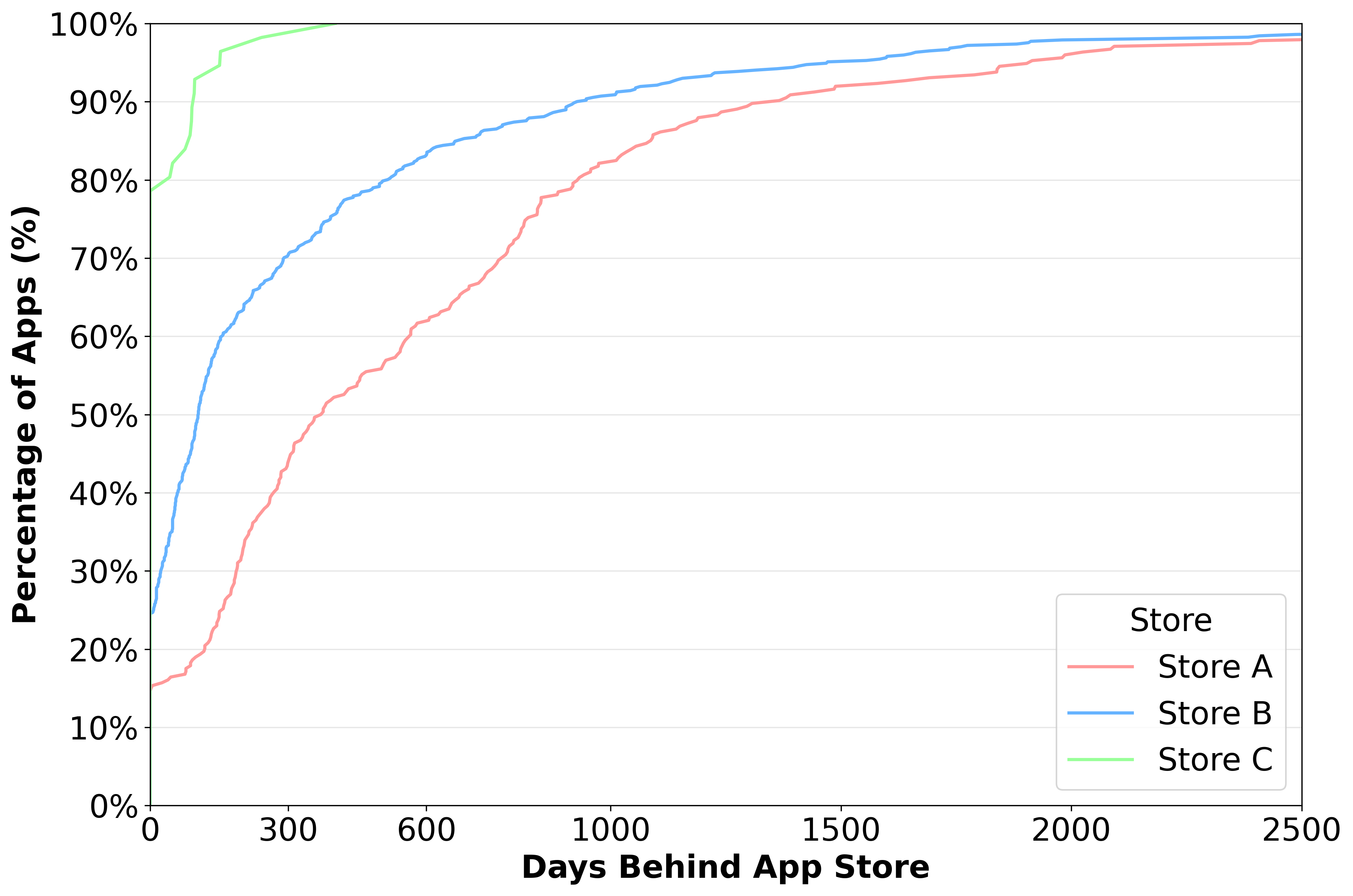}
    \caption{CDF of the delay (in days) relative to the App Store version for apps that remained unupdated in Iranian stores.}
    \label{fig:days_behind}
\end{figure}

\noindent \paratitle{Never-Updated Apps}
79.9\%, 79.1\%, and 59.0\% of apps in \sibapp, \iapps, and \appstar, respectively, have not been updated during our analysis period. Figure~\ref{fig:days_behind} illustrates the cumulative distribution of the number of days by which these apps lag behind the latest version available on the App Store. 80\% of apps in \appstar that have not been updated are indeed the latest version. In contrast, \sibapp and \iapps exhibit fewer updates, with more than 50\% of such apps being more than 4 and 13 months behind the current App Store version, respectively. In both stores, 10\% of such apps are more than 4 years behind the current version in the App Store. These delays may introduce critical security issues for their users.

\begin{figure}[tb]
    \centering
    \includegraphics[width=.9\linewidth]{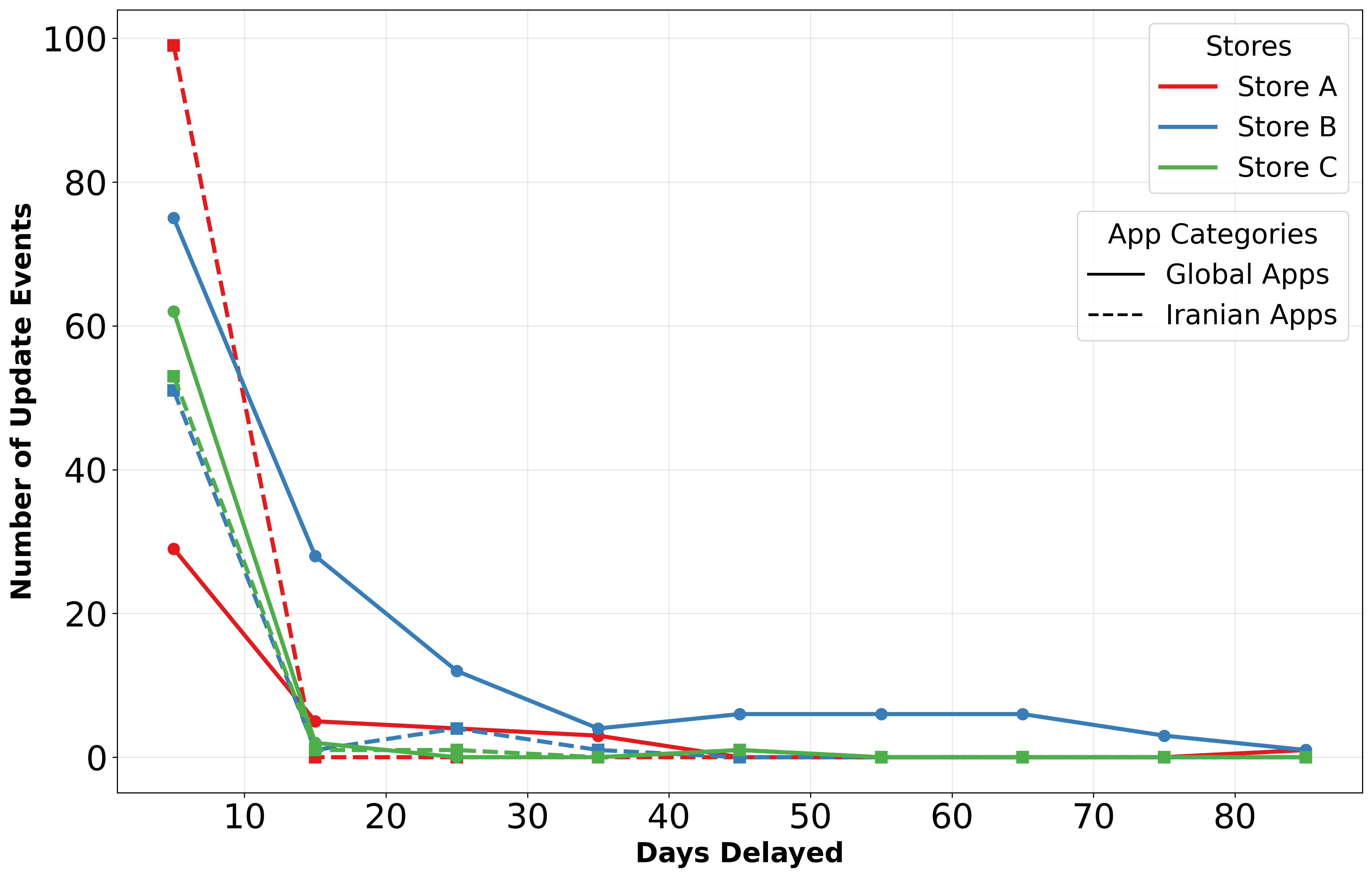}
    \caption{The number of update events versus delay (in days) for global and Iranian apps. Most updates occur within a week, with global apps experiencing longer delays, especially in \iapps.}
    \label{fig:update_delay}
\end{figure}

\paratitle{Update Delay}
We recorded total number of \APPUPDATES update events during our analysis. Figure~\ref{fig:update_delay} presents the distribution of the number of updated apps and days of delay associated with them. The majority of update events experienced delays of $<7$ days. This trend is present both in Iranian and global apps. The longest update delays were seen in WPS Office, Pou, and WhatsApp Messenger, with delays exceeding 70 days.

\subsection{Application Downloads}

Figure~\ref{fig:download-category} presents the breakdown of the number of downloads in each category in \sibapp and \iapps (\appstar does not provide download statistics). Finance and Travel are the most popular categories in these stores, each containing over a million downloads. Finance apps encompass banking services as well as money transfer platforms such as Asan Pardakht, which is a prominent platform for money transfers in Iran. International money transfer services, such as PayPal, do not operate in Iran due to sanctions. The Travel category includes internet-based taxi services such as Snapp and Tapsi. Snapp functions as a multi-purpose platform similar to Uber, allowing users to request rides, order food, and purchase travel packages.
\par Most categories are predominantly populated by Iranian apps in terms of download volume. Categories such as Finance, Travel, Utilities, Navigation, and Shopping primarily feature Iranian apps.
\par Conversely, specific categories are dominated by global apps. The Games category comprises mainly international titles, as \iapps stores host many paid global games available through the App Store. The education category is dominated by global apps, particularly language-learning apps such as Duolingo and Mondly. The Photography category also mainly involves global apps, given the absence of advanced Iranian photography apps.

The top ten Iranian and global apps are illustrated in Table~\ref{tab:popular_iran} and Table~\ref{tab:popular_global} in Appendix~B. Snapp and Snapp Driver are among the most downloaded apps in~\sibapp and~\iapps, offering a variety of services, including food ordering, hotel and flight bookings, and more. Digikala is the leading online shopping platform in Iran, serving as an alternative to Amazon. Iranians can purchase devices, appliances, and even groceries through this platform. Divar ranks among the top ten most popular apps in both stores, serving as the most prevalent platform for selling and advertising second-hand merchandise in Iran. 
\par InShot and Spotify are the most globally downloaded apps in \sibapp and \iapps. The Spotify version hosted on these stores is cracked, enabling Iranians to access the service without purchasing a subscription. This highlights the widespread popularity of Spotify in Iran, despite the presence of local music services such as Melodify. Waze also ranks among the top ten in \iapps, despite being blocked in Iran.
Photography remains the leading category among the top ten most popular global apps. All of these photography apps have been cracked, allowing users to access premium features at no cost.

\subsection{Applications Analysis}
\paratitle{Global Application Prices}
\begin{figure}[tb]
    \centering
    \includegraphics[width=1\linewidth]{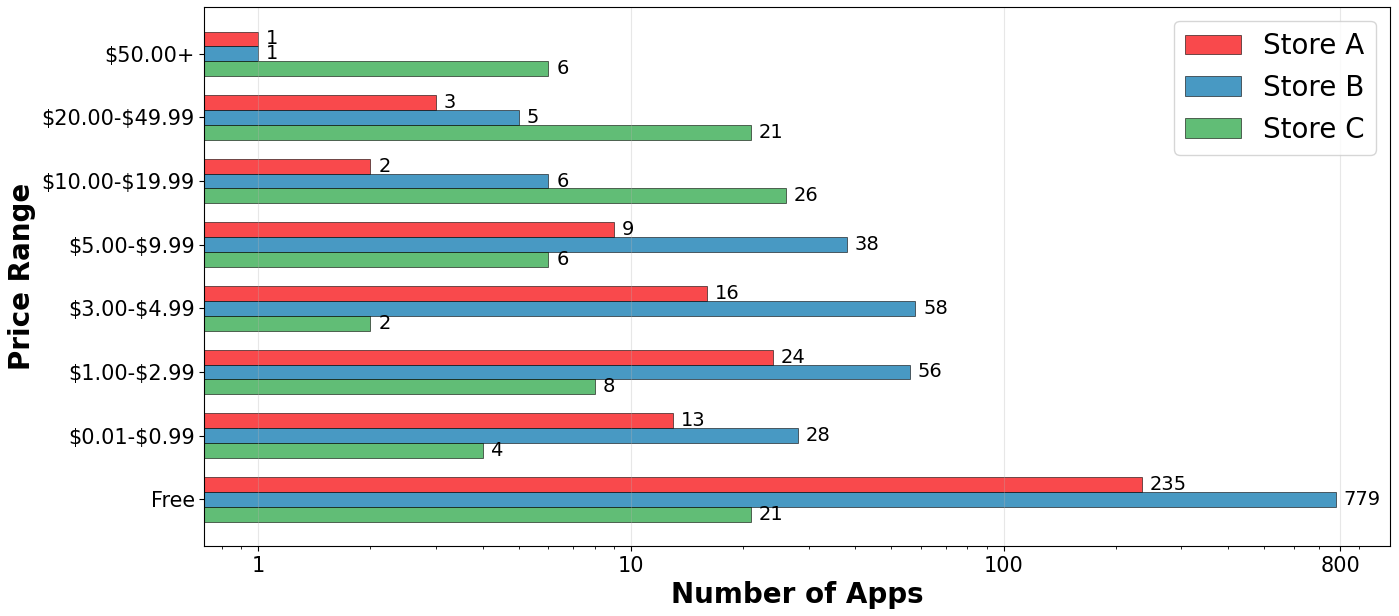}
    \caption{Price distribution of global apps.}
    \label{fig:price-dist}
\end{figure}
\begin{figure*}[tbh]
    \includegraphics[width=\linewidth]{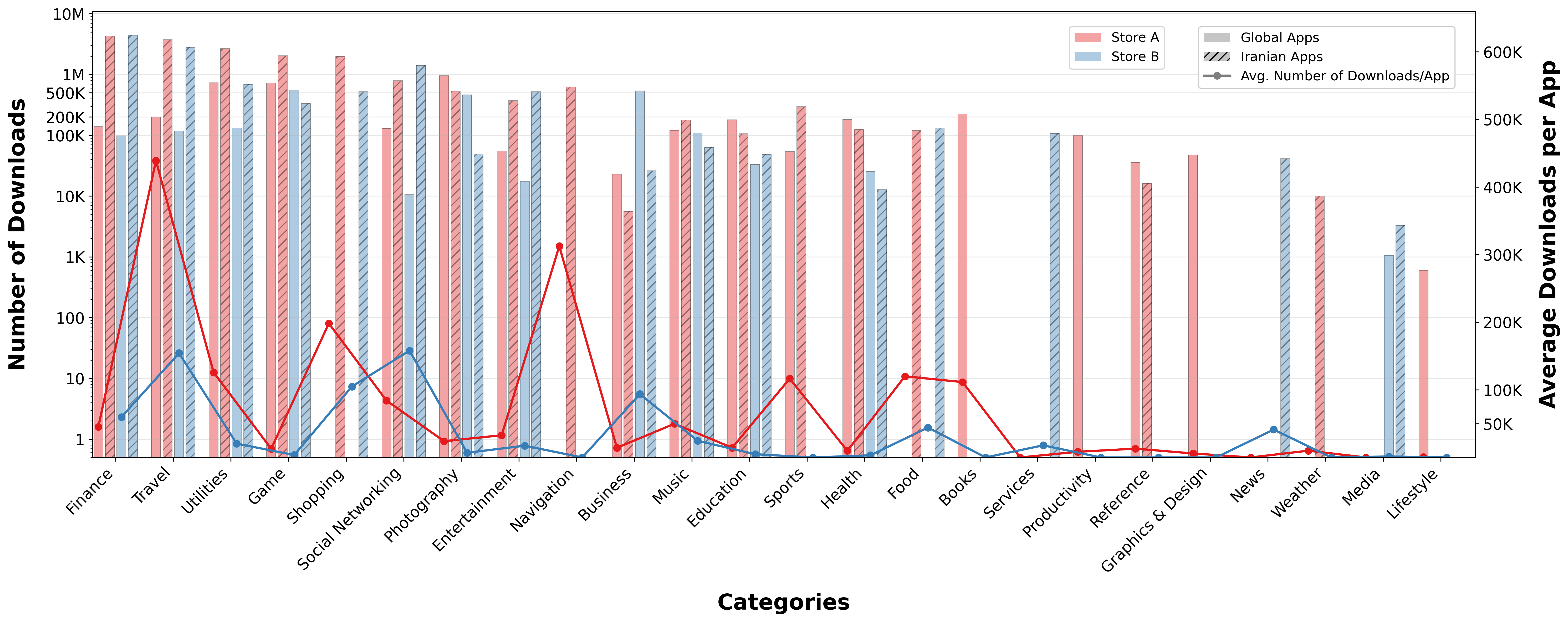}
    \captionof{figure}{Comparison of total and average app downloads across categories for \sibapp and \iapps, distinguishing between apps also available on the App Store and those exclusive to Iranian stores. Significant differences in user engagement and app distribution by category are highlighted. \vspace{-2em}}
    \label{fig:download-category}
    \end{figure*}

Figure~\ref{fig:price-dist} provides a breakdown of the prices of paid apps available on the three stores. Upon inspection, the distribution of application pricing across three app stores revealed clear variations in pricing strategies and market positioning. Free apps dominate the other pricing tiers across all stores. This suggests challenges in accessing the App Store in Iran owing to sanctions and censorship. These free apps include a significant number of cracked versions of popular apps such as Spotify, Duolingo, and PicsArt. This also indicates difficulties in making in-app purchases in Iran due to financial sanctions.
\par Mid-range price brackets (\$1.00-\$2.99 and \$3.00-\$4.99) encompass the majority of paid apps across all platforms. Most apps in this category consist of games. This suggests that the most popular paid games on the App Store fall within this specific price range.
Within the \$20-\$50 range, we identified premium video editing apps like LumaFusion and PhotoLab. AppStar offers 20 high-quality apps in this range, including premium music creation tools such as Drum School and ARP ODYSSEi, as well as professional photo backup apps such as Simple Transfer Pro. These options provide great features for enthusiasts and professionals alike.
Finally, in the \$50.00+ category, \appstar led with 6 apps, followed by \sibapp and \iapps with just 1 app. This further highlights \appstar’s focus on a premium market segment, contrasting sharply with Stores A and B’s emphasis on availability. 
\par Overall, the pricing distribution reflects distinct strategies among the three app stores: \iapps prioritizes affordability and accessibility with a large number of free and low-cost apps, \appstar targets premium users with a higher concentration of expensive apps, and \sibapp adopts a more conservative and balanced approach, focusing on free and moderately priced apps. These differences highlight the importance of aligning pricing strategies with target audiences to remain competitive in a diverse app marketplace.

\par We conducted a revenue loss analysis for \sibapp and \iapps, which provided download counts. This analysis is based on U.S. regional app prices, as the App Store does not provide services to Iranians. The only means of accessing these apps via the official App Store is to purchase gift cards from exchanges, which are generally available for the US store. Under a one-to-one substitution assumption (\ie every pirated download would have been a legitimate purchase), our findings indicate that \sibapp's total potential revenue loss is approximately  $\sim\$3,087,350$, with the \textit{Simple Transfer} app as highest revenue-losing app at  $\sim\$339,000$. \iapps's total potential revenue loss is $\sim\$2,170,653$, with \textit{Minecraft} app as the top revenue-losing app at $\sim\$335,359$. In total, this yields a combined upper-bound estimate of approximately \$5.26M across both stores. These estimates do not include subscription revenue lost due to cracked apps.

\par Previous work on piracy economics research has shown that a one-to-one substitution rate overstates actual losses, as not every pirated copy would have resulted in a legitimate purchase~\cite{gao2010, smith_telang2012}. Empirical studies report substitution rates ranging from approximately 10\% to 40\%, depending on the content type and market: Rob and Waldfogel~\cite{rob_waldfogel2006} found that music piracy displaced approximately 10\% of purchases; Peitz and Waelbroeck~\cite{peitz_waelbroeck2004} estimated a 20\% reduction in music sales due to file sharing; Volckmann~\cite{volckmann2024} measured a 19\% average revenue loss from DRM circumvention in PC games; and a European Commission study~\cite{ecorys2015} found displacement rates as high as 40\% for recent blockbuster films. While none of these studies directly map to our target domain, they present a range of possible revenue loss of \$525,800 to \$2,103,201, underscoring the substantial economic impact of app piracy in these stores.

\paratitle{Cracked Apps}
Each of the three third-party stores offers many subscription-based apps that have been altered, allowing users to access them for free or at heavily discounted prices in Iranian rial currency. Our analysis identified \totalCracked apps across Iranian stores. The most common category among these compromised apps is Educational apps, especially language learning apps such as Duolingo and Mondly. The list also includes popular apps such as Spotify and Calm.

\paratitle{Paid Iranian Apps}
\sibapp possesses several paid apps in Iranian currency, specifically \numOfAllPaidStoreA. These apps are categorized into three groups:
(1)~Apps that are Paid in both the App Store and \sibapp, (2)~Free in the App Store that are paid in \sibapp, and (3)~Paid Iranian apps that don't exist in the App Store. 
\par The first category comprises seven apps: five are photography apps, including cracked versions of LumaFusion and ProCamera, and three others. The remaining two are Simple Transfer Pro and Pocket Yoga.
The second category comprises three apps: FiLMiC Pro, Infuse, and Artstudio Pro. The annual subscription for FiLMiC Pro costs \$40 and is offered at 15,000 IRR (at the time of writing, each \$1 is equivalent to 1,500,000~IRR). The lifetime subscription for Infuse is priced at \$90 and is available for 5,000 IRR.
The final category comprises ten apps. This compilation includes both Iranian apps and those removed from the App Store, such as FIFA Street 2 and God of War: Chains of Olympus. The remaining items consist of Iranian games and an app developed by a Persian developer for the illegal downloading of content from YouTube and Spotify.

\section{App Analysis}
To assess the extent of binary modification in third-party stores, we performed a multi-layered analysis combining static binary comparison, tracker identification, and runtime traffic monitoring. We used MobSF~\cite{mobsf} to automate static analysis of each IPA and extracted embedded URLs using the \texttt{strings} utility and enumerated library dependencies with \texttt{otool}~\cite{otool}. We classify a file as \emph{injected} if its relative path appears in the Iranian-store version but not in the corresponding App Store version. For files present in both versions, we additionally verified integrity via SHA-256 hash comparison.

\paratitle{Analysis Coverage}
Of the \totalNumberOfApps unique apps in our dataset, \totalPaid are paid apps and could not be obtained for comparison, as they require purchase and their binaries are protected by FairPlay DRM.
We successfully analyzed \analyzedApps app--store pairs via MobSF, compared file listings for \fileDiffApps pairs, binary security features for \binaryCompApps pairs, App Transport Security (ATS) configurations for \atsCompApps pairs, and embedded third-party tracker domains for all \analyzedApps pairs. For runtime analysis, we installed each app version on a physical iPhone X running iOS 16.7.15 and captured network traffic through a MITM proxy to compare traffic destinations and plaintext exposure across store variants.

\paratitle{Injected Files and Crack Tools}
Our comparison of IPA file listings between App Store and Iranian store versions revealed significant structural differences. Of the \fileDiffApps apps with file-level differences, \crackToolApps (67.9\%) contained at least one known crack tool or piracy-related artifact. Table~\ref{tab:missing_files_top10} lists the most frequently injected files.

\begin{table}[tb]
\centering
\caption{Top 10 injected files found in Iranian store IPAs but absent from the corresponding App Store version.}
\label{tab:missing_files_top10}
\footnotesize\begin{tabular}{lp{4.8cm}r}
\toprule
File Name  & Description & Apps \\
\midrule
mobileprovision  & Ad-hoc signing provisioning profile & 489 \\
Cydia Substrate  & Runtime ObjC method-hooking framework & 186 \\
SignedByEsign  & Marker artifact left by Esign platform & 179 \\
CydiaSubstrate.h     & Cydia Substrate header (re-signing artifact) & 142 \\
iGameGod  & In-game memory cheat tool & 68 \\
libsubstrate.dylib   & Cydia Substrate shared library & 81 \\
cyan.entitlements     & Entitlements injected by iGameGod & 51 \\
libswiftIU.dylib   & iGameGod Swift UI library & 48 \\
CrackerXI      & App Store binary decryption/dump tool & 28 \\
AlertTheos.dylib        & Theos-based code injection library & 20 \\
\bottomrule
\end{tabular}
\end{table}

\paratitle{Embedded Links Analysis}
Table~\ref{tab:missing_links_top10} presents the top 10 domains embedded in Iranian store IPAs but absent from the App Store version. The majority of added links belong to re-signing endpoints or social media accounts of the application crackers.

\begin{table}[tb]
\caption{Top 10 extra domains embedded in Iranian store IPAs. }
\label{tab:missing_links_top10}
\centering
\footnotesize\begin{tabular}{llr}
\toprule
Domain  & Description & Apps \\
\midrule
yyyue.xyz  & Esign signing platform endpoint & 190 \\
t.me  & Cracker Telegram channel & 177 \\
twitter.com & Cracker social media links & 105 \\
iosgods.com  & Cracked app distribution platform & 92 \\
github.com & Cracking tool repositories & 87 \\
saurik.com & Cydia Substrate developer homepage & 75 \\
igamegod.app & Game cheat tool server & 74 \\
startappservice.com & Ad network & 74 \\
startapp.com & Ad network & 72 \\
\bottomrule
\end{tabular}
\end{table}

\paratitle{App Transport Security}
We compared ATS configurations in \texttt{Info.plist} across \atsCompApps app pairs. By default, ATS requires HTTPS with TLS 1.2 or higher, forward secrecy, and strong ciphers, though apps can declare exceptions that relax these requirements for specific domains. Of these pairs, \atsDiffApps showed ATS configuration differences between their Iranian and App Store versions, and \atsWeakerApps of those adopted strictly weaker settings in the Iranian version: they added domain exceptions, enabled \texttt{NSAllows\-Arbitrary\-Loads} globally, lowered \texttt{NSMinimum\-TLS\-Version}, or disabled forward secrecy requirements that the App Store version enforced. These weakened configurations allow insecure or downgraded connections to domains the original developer chose to protect with strong TLS.

\paratitle{Permission Analysis}
Of the apps compared, \permDiffApps exhibited permission differences between Iranian and App Store versions, all in \sibapp. These apps requested additional permissions beyond those declared in the App Store version: \textit{Subway Surfers} gained calendar;
\textit{Shadow Fight 2} gained camera and motion;
and \textit{Collect} app gained location permission.
While most permission changes are minor, any additional permission not present in the original binary represents an undisclosed capability introduced without user consent.

\paratitle{Runtime Behavioral Analysis}
To complement our static analysis, we captured runtime traffic for available IPAs. Each version from each store was installed on a physical iPhone, and its network traffic was recorded through a MITM proxy for 20~seconds per launch, allowing us to compare traffic destinations and plaintext exposure across store variants. In total, we captured \totalFlows network flows across \appsWithFlowsAppStore App Store apps, \appsWithFlowsIApps \iapps apps, and \appsWithFlowsSibApp \sibapp apps. 

Of the \totalFlows captured flows, \httpsFlows (99.1\%) used HTTPS, and only \plaintextFlows ({\plaintextPct}) were sent over plaintext HTTP, affecting \plaintextApps unique apps. Importantly, these plaintext flows were not introduced by the repackaging process: they appeared in both the App Store and Iranian store versions of the same apps, indicating that they originate from the app's own code (\eg legacy analytics endpoints, image CDNs) rather than from injected libraries. This confirms that iOS ATS enforcement at the OS level effectively prevents the cracking process from downgrading transport-layer security.

We also compared the set of domains each app contacted across stores. For \flowPairsCompared app-version pairs where we captured traffic from both the App Store and at least one Iranian store, we identified domains appearing exclusively in the Iranian store version. Table~\ref{tab:runtime_domains} lists the most prominent such domains. The Apple sandbox endpoints indicate that repackaged apps connect to Apple's sandbox StoreKit environment rather than production payment infrastructure---a direct consequence of the re-signing process stripping FairPlay DRM. The non-Apple domains confirm that injected ad SDKs and game modification tools are active at runtime rather than dormant.

\begin{table}[tb]
\centering
\caption{Top domains contacted at runtime exclusively in Iranian store versions (absent from App Store versions).}
\label{tab:runtime_domains}
\small
\footnotesize\begin{tabular}{p{3.55cm}cl}

\toprule
Domain & Count & Context \\
\midrule
sandbox.itunes.apple.com & 395 & App Resigning \\
amp-api.sandbox.apple.com & 219 & App Resigning \\
mzstorekit-sb.itunes.apple.com & 176 & App Resigning \\
file.ipaomtk.com & 40 & Cracked app distributor \\
www.iosgods.com & 10 & Cracked app distributor \\
api.mixpanel.com & 9 & Injected analytics SDK \\
ms4.applovin.com & 14 & Injected ad network SDK \\
igamegod.app & 5 & Game cheat tool server \\
\bottomrule
\end{tabular}
\end{table}

\paratitle{Injected Library Runtime Behavior}
Beyond traffic analysis, our static findings provide evidence of behavioral risks from the injected libraries. Table~\ref{tab:injected_lib_behavior} summarizes the runtime capabilities of the most prevalent ones.

\begin{table}[tb]
\centering
\caption{Runtime capabilities of injected libraries.}
\label{tab:injected_lib_behavior}
\small
\footnotesize\begin{tabular}{lcp{5cm}}
\toprule
Library & Apps & Runtime Capability \\
\midrule
Cydia Substrate & 186 & Hooks ObjC methods via \texttt{MSHookMessageEx}; bypasses payment/auth handlers \\
iGameGod & 68 & Modifies game memory at runtime; contacts \texttt{igamegod.app} at launch \\
AlertTheos.dylib & 20 & Injects code into app process at launch via Theos framework \\
SatellaJailed & 7 & Intercepts StoreKit to return fabricated purchase receipts \\
\bottomrule
\end{tabular}
\end{table}

Collectively, these libraries introduce capabilities that extend substantially beyond in-app purchase circumvention. A Substrate hook can redirect API responses, capture credentials, or disable certificate pinning without any visible change to the user interface. Combined with the runtime connections to cracking infrastructure and ad networks documented above, the injected libraries represent a compounding risk to user privacy and security.

\section{Discussion}
The unique dynamic created because of sanctions and censorship, and the use of third-party app stores, poses unique challenges absent from traditional mobile platforms. We break these challenges into four key categories:
\begin{itemize}[noitemsep,topsep=0pt]
    \item \textit{Outdated Apps:} While outdated apps are typically considered a non-issue in the iOS ecosystem due to automated updates enforced by the App Store, our results highlight them as a key challenge in third-party app store usage. Updates are often delayed or absent, causing users to run outdated versions of the applications for weeks or months, even when known bugs and security issues have been addressed by the developer.

    \item \textit{Lack of review and weak integrity checks:} Apps installed outside the App Store do not pass through Apple's review process, creating opportunities for tampering, repackaging, and other unauthorized modifications. The presence of injected libraries such as Cydia Substrate, iGameGod, AlertTheos.dylib, and SatellaJailed indicates that many of these apps have been altered after release. These tools are designed for runtime hooking or purchase bypass rather than passive inclusion; for example, Cydia Substrate can intercept Objective-C method calls and modify behaviors related to authentication and payment verification.
 Combined with unauthorized ad and tracking SDKs, such modifications expose users to behavioral profiling and possible data exfiltration without their knowledge.

\item \textit{Dataloss and Instability Caused by Revocations:} Third-party stores typically depend on the illicit use of Apple developer certificates, which are frequently detected and revoked. When Apple revokes these certificates, the affected apps immediately stop functioning.
This poses a serious side effect: if an app stores data locally, users may lose that data, which could include progress, notes, files, saved content, and other information that was never backed up.

\item \textit{Lack of Access to Free Apps:}
Although General License D-2 allows Apple to provide fee-based and no-cost services and software to Iranians users, existing obstacles for creating Apple ID has prevented equitable access to this population. These difficulties have led to the proliferation of third-party app stores and creation of underground markets for Apple IDs at huge security and privacy risks to users. Correct implementation of these policies can have a significant effect in decreasing user's reliance on third-party markets.

\end{itemize}

\section{Related Work}
\paratitle{App Stores}
 A multitude of previous studies \cite{same_app,ios_decision,removed_ios,appstore_software_analysis,android_trust,google_play_ml,hey_you,ios_privacy_labels} have performed measurement and analysis on Apple and Android app stores. Fuqi Lin \etal \cite{removed_ios} have conducted longitudinal studies to examine the patterns and characteristics of mobile apps that are gradually removed from the App Stores. DroidRanger~\cite{hey_you} has been developed to detect malware in alternative Android markets, which operate using modified or unofficial versions of the Google Play store. This system aims to enhance security by identifying malicious apps that may not be present in the official store. Gian Luca Scoccia \etal have conducted empirical investigations \cite{ios_privacy_labels} into the privacy labels associated with apps published on the App Store, analyzing how these labels communicate privacy practices to users and assessing their accuracy and consistency across different apps. Overall, this body of related work contributes to a comprehensive understanding of app lifecycle, security, and privacy considerations within the mobile app ecosystem. However, compared with the Android ecosystem, less work has been undertaken in the iOS ecosystem, particularly on the App Store and third-party stores~\cite{ios_overlooked}; our focus is directed towards that area.\\
\paratitle{Mobile Application Security} Prior studies \cite{pios,automated_binary_crypto,crios,tapping_ipa,android_repackaged,hey_you} have systematically analyzed various aspects related to the security of mobile apps. Steven Selden \etal have developed AppTap~\cite{tapping_ipa}, which decrypts iOS binaries by utilizing an ARM-based Mac, allowing researchers to perform dynamic analysis on these libraries to identify potential vulnerabilities. CRiOS~\cite{crios} concentrates on the static analysis of iOS app packages (IPAs) downloaded from the Official App Store. This process involves decrypting the IPAs on a jailbroken iPhone to examine the app's code and behavior in greater detail. In Android security, DroidMoss~\cite{android_repackaged} is a specialized tool for detecting repackaging of Android apps. It collects packages from Google Play and various third-party app stores, then compares these packages using a similarity scoring algorithm to identify unauthorized modifications. PiOS~\cite{pios} is aimed at detecting privacy leaks in iOS apps by analyzing class-dump files, which reveal the internal structure of app binaries and help identify sensitive data flows that may compromise user privacy.\\
\paratitle{Iran Digital Ecosystem}
 Previous research \cite{iran_appstores,mobile_health_iran,banking_iran,Yalcintas_Alizadeh_2020,mobile_downloads_cafe,taxi_iran,entrepre_iran} has explored the Iranian digital ecosystem from multiple angles.  Jafari~\etal~\cite{mobile_downloads_cafe} analyzes the aesthetic components that influence app downloads, specifically 
 examining a third-party Android store known as Cafebaazar \cite{cafebazaar} in Iran, and investigates how visual and design elements impact user engagement. Roshandel~\etal~\cite{taxi_iran} concentrates on two major online taxi platforms operating in Iran, thoroughly analyzing their business strategies, marketing approaches, and competitive advantages in the Iranian market. Asadi~\etal~\cite{iran_appstores} examined the influence of online reviews on paid mobile apps, specifically assessing how user feedback and ratings affect app popularity and user trust. Fadaeizadeh~\etal~\cite{mobile_health_iran}  evaluated the top Iranian medical and health apps from the Cafebaazar App Store \cite{cafebazaar}, assessing both user and expert perspectives on content and technical performance. The findings reveal a gap between user popularity and expert quality, highlighting the lack of formal standards for app content and design in this field.
\section{Key Takeaways \& Conclusion}
In this paper, we presented the first comprehensive empirical study of unauthorized Iranian iOS app stores. Our study reveals how sanctions, censorship, and technical restrictions have collectively contributed to the emergence of unauthorized third-party app stores. Through a 5.5-month analysis of more than \totalUniqueApps applications from three major Iranian stores, we observed that these platforms fill a critical access gap while simultaneously undermining software integrity, user security, and developer revenues.

Our findings underscore several broader lessons.
First, sanctions and platform-level restrictions can unintentionally incentivize the creation of parallel digital ecosystems that operate outside traditional oversight and security frameworks. The Iranian iOS market illustrates how users are willing to adapt to maintain digital access, even when doing so exposes them to pirated or tampered software.
Second, this ecosystem reflects the deep entanglement of technical and geopolitical factors, in which policy decisions aimed at restricting access inadvertently promote piracy, undermine user safety, and complicate legitimate development efforts.
Third, the persistence of outdated, modified, or cracked binaries highlights how the lack of official access erodes software reliability and introduces new vectors for surveillance and exploitation.

From a policy perspective, our results suggest that blanket sanctions and restrictive compliance mechanisms should be re-evaluated in light of their digital spillover effects. Efforts to safeguard users and developers must account for informal adaptation strategies that flourish in constrained environments. For platform operators, the Iranian case serves as a cautionary model of how tightly coupled ecosystems, such as iOS, can fracture under geopolitical pressure.

The Iranian third-party iOS app ecosystem is not merely a regional anomaly. It is a case study in how digital markets evolve under constraints. Understanding these unintended consequences is crucial for developing technology governance and international policies that promote security, equity, and resilience in the global software ecosystem.

\bibliographystyle{ACM-Reference-Format}
\bibliography{amir,ref}

\appendix
\section{Ethics}
This work examines a sensitive and legally complex ecosystem at the intersection of international sanctions, platform policies, and state-level internet censorship. As such, we considered the ethical implications of our methodology, data collection, analysis, and dissemination, with particular attention to potential harm to users, developers, and platform operators.

\paratitle{Human Subjects and User Harm}
Our study does not involve direct interaction with human subjects, nor does it collect personal data such as names, phone numbers, Apple IDs, payment credentials, or device identifiers associated with individual users. While Iranian third-party app stores require phone numbers and device UDIDs for subscription and app installation, our measurement infrastructure did not register or retain any personally identifying information beyond what was strictly necessary to operate a single test device. We did not attempt to deanonymize users, infer user identities, or link app usage to individuals. Consequently, this research does not constitute human-subjects research under common IRB definitions.

\paratitle{Data Collection and Privacy}
We collected app metadata and IPA binaries from publicly accessible third-party app store interfaces and from Apple’s official App Store APIs. The metadata collected reflects information already visible to users of these platforms. Static analysis was performed exclusively on application binaries and did not involve intercepting live user traffic or monitoring user behavior. When comparing third-party IPAs to official App Store versions, we focused on structural and library-level differences rather than any user-generated content.
To minimize risk, we do not release raw IPA files or decrypted binaries publicly. Any dataset shared with other researchers will be subject to access controls and fair-use agreements.

\paratitle{Legality and Policy Compliance}
We acknowledge that some of the observed practices in Iranian third-party iOS app stores  (\eg unauthorized redistribution, cracking of paid apps, and misuse of Apple developer certificates) violate Apple’s developer policies and may infringe on intellectual property rights. Our research does not endorse, facilitate, or operationalize these practices. Instead, our goal is to measure and characterize an already-existing ecosystem to understand its scale, risks, and broader implications.

Our measurement activities were conducted solely for academic research purposes. We did not distribute cracked apps, provide instructions for bypassing safeguards, or assist users in evading sanctions or platform restrictions. We relied on read-only analysis and avoided actions that could materially strengthen or sustain these underground markets.

\paratitle{Dual-Use and Misuse Considerations}
This work necessarily describes the technical mechanisms underlying third-party app stores, including the misuse of ad hoc and enterprise signing. To mitigate dual-use risks, we intentionally frame these mechanisms at the conceptual and analytical levels, avoiding step-by-step instructions, tool releases, or operational guidance that could lower the barrier to abuse. The intent of this work is to inform platform operators, policymakers, and security researchers about systemic weaknesses and unintended consequences, not to enable replication.

\paratitle{Harm–Benefit Tradeoff}
We carefully weighed the potential harms of exposing weaknesses in the iOS distribution model against the benefits of transparency. We believe the benefits outweigh the risks for several reasons. First, the ecosystem we study is already widely used by millions of users and operates at scale. Second, the lack of visibility into this ecosystem exacerbates security and privacy risks for users who rely on it out of necessity. Third, understanding how sanctions and access restrictions shape software ecosystems is critical for developing more equitable and secure platform policies.

\paratitle{Responsible Disclosure and Broader Impact}
Where our findings reveal systemic risks, such as outdated binaries, unauthorized tracking libraries, and widespread certificate misuse, we present them in aggregate. Our goal is to encourage informed discussion among platform operators, regulators, and researchers about safer alternatives for app access under geopolitical constraints. We deliberately avoid naming third-party stores.

Due to ethical, legal, and safety considerations described in the above section, decrypted IPA binaries, certificates, or store credentials are intentionally not released. These artifacts will be available to
other researchers, upon request, subject to fair use provisions
and appropriate data use agreements.

\section{Additional Tables and Figures}

\begin{table*}
\centering
\caption{Top 10 popular Iranian apps based on the number of downloads.}
\resizebox{.8\textwidth}{!}{
\begin{tabular}{@{}cllll|llll@{}}
\toprule
& \multicolumn{4}{c}{\textbf{\sibapp}} & \multicolumn{4}{c}{\textbf{\iapps}} \\ \cmidrule(l){2-9}
\multicolumn{1}{l}{\textbf{Ranking}} 
& \multicolumn{1}{l}{\textbf{App}} 
& \multicolumn{1}{l}{\textbf{Category}} 
& \multicolumn{1}{l}{\textbf{Downloads}}
& \textbf{Description} 
& \multicolumn{1}{l}{\textbf{App}} 
& \multicolumn{1}{l}{\textbf{Category}} 
& \multicolumn{1}{l}{\textbf{Downloads}}
& \textbf{Description} \\ \toprule
1  & Snapp        & Travel            & 1,730,000 & Ride Share & Snapp Driver   & Travel            & 718,677 & Ride Share \\ \midrule
2  & Digikala     & Shopping          & 1,090,000 & Amazon Alternative & Snapp          & Travel            & 590,694 & Ride Share \\ \midrule
3  & MyIrancell   & Utilities         & 1,010,000 & ISP Application & Bank Mellat    & Finance           & 583,682  & Bank App \\ \midrule
4  & TAPSI        & Travel            & 905,000 & Ride Share & Eitaa          & Social Networking & 559,206 & WhatsApp Alternative \\ \midrule
5  & MyMCI        & Utilities         & 865,000 & ISP Application & Neshan         & Travel            & 550,735 & Navigation App \\ \midrule
6  & Asan Pardakht& Finance           & 715,000 & PayPal Alternative & Asan Pardakht  & Finance           & 542,433 & PayPal Alternative \\ \midrule
7  & Divar        & Utilities         & 700,000 & Craigslist Alternative & Divar          & Business          & 527,727 & Craigslist Alternative \\ \midrule
8  & Sheypoor     & Shopping          & 505,000 & Craigslist Alternative & Baam           & Finance           & 493,927 & PayPal Alternative \\ \midrule
9  & Filimo       & Entertainment     & 485,000 & Netflix Alternative & Bank Saderat   & Finance           & 440,867 & Bank App \\ \midrule
10 & Mellat bank  & Finance           & 460,000 & Bank App & Balad          & Travel            & 421,235 & Navigation App \\
\bottomrule
\label{tab:popular_iran}
\end{tabular}
}
\\
\vspace{-2em}
\end{table*}
\begin{table*}
\footnotesize
\centering
\caption{Top 10 popular global apps in each store based on the number of downloads.}
\begin{tabular}{@{}clll|lll@{}}
\toprule
                              & \multicolumn{3}{c}{\textbf{\sibapp}}                                           & \multicolumn{3}{c}{\textbf{\iapps}}                                           \\ \cmidrule(l){2-7} 
\multicolumn{1}{l}{\textbf{Ranking}} & \multicolumn{1}{l}{\textbf{App}} & \multicolumn{1}{l}{\textbf{Category}} & \textbf{Downloads} & \multicolumn{1}{l}{\textbf{App}} & \multicolumn{1}{l}{\textbf{Category}} & \textbf{Downloads} \\ \toprule
1                             & \multicolumn{1}{l}{InShot}    & \multicolumn{1}{l}{Photography}         & 130,000             & \multicolumn{1}{l}{Spotify}    & \multicolumn{1}{l}{Music}         & 97,604            \\ \midrule
2                             & \multicolumn{1}{l}{Spotify}    & \multicolumn{1}{l}{Music}         & 120,000              & \multicolumn{1}{l}{InShot}    & \multicolumn{1}{l}{Photography}         & 94,970             \\ \midrule
3                             & \multicolumn{1}{l}{Picsart}    & \multicolumn{1}{l}{Photography}         & 120,000             & \multicolumn{1}{l}{Waze}    & \multicolumn{1}{l}{Navigation}         & 82,794             \\ \midrule
4                             & \multicolumn{1}{l}{Mondly}    & \multicolumn{1}{l}{Education}         & 110,000              & \multicolumn{1}{l}{CapCut}    & \multicolumn{1}{l}{Photography}         & 74,780             \\ \midrule
5                             & \multicolumn{1}{l}{Calm}    & \multicolumn{1}{l}{Health}         & 65,000              & \multicolumn{1}{l}{SHAREit Pro }    & \multicolumn{1}{l}{Utilities}         & 72,067             \\ \midrule
6                             & \multicolumn{1}{l}{Game of Cards}    & \multicolumn{1}{l}{Game}         & 64,000             & \multicolumn{1}{l}{ WhatsApp Watusi}    & \multicolumn{1}{l}{Social Networking}         & 61,534              \\ \midrule
7                             & \multicolumn{1}{l}{Facetune AI}    & \multicolumn{1}{l}{Photography}         & 57,000             & \multicolumn{1}{l}{PicsArt}    & \multicolumn{1}{l}{Photography}         & 50,403             \\ \midrule
8                             & \multicolumn{1}{l}{Focos}    & \multicolumn{1}{l}{Photography}         & 45,000             & \multicolumn{1}{l}{Minecraft}    & \multicolumn{1}{l}{Game}         & 49,729             \\ \midrule
9                             & \multicolumn{1}{l}{Bazaart}    & \multicolumn{1}{l}{Photography}         & 43,000             & \multicolumn{1}{l}{Pou}    & \multicolumn{1}{l}{Game}         & 33,898             \\ \midrule
10                             & \multicolumn{1}{l}{ProCamera}    & \multicolumn{1}{l}{Photography}         & 42,000             & \multicolumn{1}{l}{LIMBO}    & \multicolumn{1}{l}{Game}         & 31,647             \\
\bottomrule
\label{tab:popular_global}
\end{tabular}
\end{table*}

\end{document}